\documentclass[aps,prb,twocolumn]{revtex4-2}

\usepackage{amsfonts}
\usepackage{amsmath}
\usepackage{graphicx}
\usepackage{comment}
\usepackage{physics}
\usepackage{dsfont}
\usepackage{bbm}

\newcommand\identity{\mathbbm{1}}
\newcommand{\beq}{\begin{equation}}
\newcommand{\eeq}{\end{equation}}
\newcommand{\beqnn}{\begin{equation*}}
\newcommand{\eeqnn}{\end{equation*}}
\newcommand{\bea}{\begin{eqnarray}}
\newcommand{\e}{\text{e}}
\newcommand{\eea}{\end{eqnarray}}
\newcommand{\beann}{\begin{eqnarray*}}
\newcommand{\eeann}{\end{eqnarray*}}
\newcommand{\bes} {\begin{subequations}}
\newcommand{\ees} {\end{subequations}}

\usepackage{xcolor}
\newcommand\note[1]{}

\newcommand\n{{\bf n}}

\begin{document}
\title{A quantum Monte Carlo algorithm for Bose-Hubbard models on arbitrary graphs}
\author{Emre Akaturk}
\affiliation{Information Sciences Institute, University of Southern California, Marina del Rey, California 90292, USA}
\author{Itay Hen}
\affiliation{Information Sciences Institute, University of Southern California, Marina del Rey, California 90292, USA}
\affiliation{Department of Physics and Astronomy and Center for Quantum Information Science \& Technology, University of Southern California, Los Angeles, California 90089, USA}
\email{itayhen@isi.edu}

\begin{abstract}
\noindent We propose a quantum Monte Carlo algorithm capable of simulating the Bose-Hubbard model on arbitrary graphs, obviating the need for devising lattice-specific updates for different input graphs.  We show that with our method, which is based on the recently introduced Permutation Matrix Representation Quantum Monte Carlo [Gupta, Albash and Hen, J. Stat. Mech. (2020) 073105],  the problem of adapting the simulation to a given geometry amounts to generating a cycle basis for the graph on which the model is defined, a procedure that can be carried out efficiently and and in an automated manner. To showcase the versatility of our approach, we provide simulation results for Bose-Hubbard models defined on two-dimensional lattices as well as on a number of random graphs.
\end{abstract}

\maketitle

\section{Introduction}

\note{DEFINE \\identity}

\note{Computation/QMC background}

\note{Motivation: No general QMC, We do any graph, not just physical systems using PMR-QMC}

The Bose-Hubbard (BH) model, one of the pillars of condensed matter physics, is the go-to model for a large variety of physical phenomena, from Mott-Insulator-to-superfluid transitions to bosonic atoms in optical lattices. Similar to many other fundamental quantum systems of importance in condensed matter physics, the BH model does not admit analytical solutions in the general case and studying it usually requires resorting to approximation techniques, as even exact-numerical methods become unfeasible with increasing system size.

The most common approach for studying the BH model is statistical Quantum Monte Carlo  (QMC) techniques~\cite{Lewenstein12,Fisher89,Jaksch05,Giamarchi08}.
QMC has been used to study the BH model throughout the years in a variety of contexts. Among these are supersolid phases~\cite{Batrouni95,Hebert01,Sengupta05,Heidarian05,Wessel05,Boninsegni05,Gan07,Bogner19}, superfluid to Mott insulator transition~\cite{Krauth91,Krauth91-2,Kisker97,Melko05,Yokoyama11} and superfluid to Bose glass transitions~\cite{Krauth91,Scalettar91,Kisker97,Sengupta07}. Other studies focus on the BH model manifested on optical lattices with confining potentials~\cite{Prokofev03,Wessel04,Wessel05-2,Pollet_2008} and extensions thereof~\cite{Sengupta05,Gan07,Gan07-2,Kawaki17,Bogner19}.

Different setups of the BH model varying in both dimension and geometry have been explored, most notably with the Stochastic Series Expansion technique~\cite{sandvik:92,sandvik:99,sandvik:05,sandvik:10}, employing different types of updates including dual vortex theory~\cite{Isakov06}, multi-site generalization~\cite{Isakov09} or directed loops~\cite{Gan07}.  Other examples include studying the model on one-dimensional lattices~\cite{Sengupta07,PhysRevA.80.033612,Kawaki17,Scalettar91,PhysRevB.46.9051,Wessel05-2,Pollet_2008}, triangular~\cite{Boninsegni05,Heidarian05,Wessel05,Melko05,Gan07} or rectangular lattices in two dimensions~\cite{Krauth91,Krauth91-2,vanOtterlo94,Batrouni95,Kisker97,Batrouni00,Hebert01,Sengupta05,Capogrosso-Sansone08,Yokoyama11,Soyler11,Ohgoe12,Bogner19} and cubic lattices in three dimensions~\cite{Prokofev03,Capogrosso-Sansone07,Anders10}. Other lattice types include a cubic lattice with a harmonic confining potential ~\cite{Kato09},  the kagome lattice~\cite{Isakov06}, the star lattice~\cite{Isakov09}, the honeycomb lattice~\cite{Gan07-2} and more~\cite{Hettiarachchilage13}. %

One notable observation from the above survey is that simulating the BH model on different lattice structures and in different dimensions with QMC often requires one to concoct specially tailored QMC updates for each such setup. In this study, we present a resolution to this obstacle by proposing a quantum Monte Carlo simulation technique that is applicable to Bose-Hubbard models defined on arbitrary input graphs, obviating the need for implementing lattice-specific update rules for each setup separately. The proposed technique may be used to simulate the BH model on any graph and in any dimension (for the first time, as far as the authors are aware).

Our approach builds on the parameter-free Trotter error-free Permutation Matrix Representation (PMR) quantum Monte Carlo technique introduced in Ref.~\cite{Gupta20} for spin systems, wherein the quantum partition function is expanded in a power series of the off-diagonal strength of the Hamiltonian, augmented with the necessary modifications that allow simulations of the Bose-Hubbard model on arbitrary graphs. 
Specifically, we show that QMC updates guaranteeing ergodicity and which also maintain detailed balance can be achieved by generating what is known as a minimal cycle basis on the BH graph~\cite{Berger04} -- the set of cycles that form a basis for all cycles on the graph~\cite{Uno14}. 

We validate our proposed algorithm by simulating the Bose-Hubbard model on regular lattices as well as on a number of  irregular graphs with up to $64$ sites and with varying numbers of particles and Hamiltonian parameters to showcase the capabilities of our technique. 

The paper is structured as follows: In Sec.~\ref{sec:methodology}, we provide an overview of the PMR quantum Monte Carlo technique, followed by the specifics of our proposed QMC algorithm adapted to simulating BH models on arbitrary graphs. Section~\ref{sec:measurements} is devoted to illustrating how a very wide variety of measurements may be carried out, including quantities such as superfluid density and the one-body density matrix. We then move on to explain the concept of minimal cycle basis and its usage in the generation of the QMC updates for the algorithm in Sec.~\ref{sec:mcb}. In Sec.~\ref{sec:results}, we present some simulation results for a number of Bose-Hubbard models defined on a variety of graphs. We summarize our work in Sec.~\ref{sec:conc} along with some conclusions and a discussion of future work. 

\section{The QMC algorithm\label{sec:methodology}}

Our proposed QMC algorithm builds on the recently introduced Permutation Matrix Representation QMC (PMR-QMC) method~\cite{Gupta20}. Below we provide a brief overview of the general methodology, which we then discuss in more detail in the context of the Bose-Hubbard model.

\subsection{Permutation matrix representation}

The basis for the PMR-QMC method begins with casting the to-be-simulated Hamiltonian $H$ in PMR form, namely, as
\begin{equation}
\label{eq:hamiltonian_pmr}
H = \sum_{j=0}^M \tilde{P_j} = \sum_{j=0}^M D_j P_j =  D_0 + \sum_{j=1}^M D_j P_j \,,
\end{equation}
where $\{ \tilde{P_j} \}_{j=0}^M$ is a set of $M+1$ distinct generalized permutation matrices~\cite{Joyner08} -- matrices that have at most one nonzero element in each row and each column. One can write each $\tilde{P_j}$ as $\tilde{P_j} = D_j P_j$ where $D_j$ is a diagonal matrix and $P_j$ is a bonafide permutation matrix. One of the permutations,  which we denote by $P_0$, can always be chosen to be $P_0 = \identity$ (the identity operation), such that the other permutation matrices have no fixed points, i.e., no nonzero diagonal elements. We refer to the basis in which the $\{ D_j \}$ matrices are diagonal as the computational basis and denote its states by $\{ \vert z \rangle \}$. The operators ${D_jP_j}$ for $j>0$ represent the `quantum dimension' of the Hamiltonian. Acting with a $D_jP_j$ matrix on a basis state $\vert z \rangle$  gives $D_jP_j \vert z \rangle = d_j(z^\prime) \vert z^\prime \rangle$ where $d_j(z^\prime)$ is a (generally complex) coefficient and $\vert z^\prime \rangle$ is a basis state $\vert z \rangle \neq \vert z^\prime \rangle$. 
We will refer to $D_0$ (the matrix multiplying $P_0$) as the `classical Hamiltonian'. The permutation matrices derived from $H$ are a subset of the permutation group wherein $P_0$ is the identity element~\cite{Gupta20}. One can show that any finite-dimensional (or countable infinite-dimensional) matrix can be written in PMR form~\cite{Gupta20}.

\subsection{The off-diagonal partition function expansion}

Having cast the Hamiltonian in PMR form, one proceeds with expanding the canonical partition function $Z = Tr[ e^{-\beta H} ]$ about its diagonal part in powers of its off-diagonal strength~\cite{Gupta20}. The expansion results in the following expression for the partition function (a detailed derivation can be found in Appendix~\ref{apx:odp} and in Ref.~\cite{Gupta20}). 
\begin{equation}
\label{eq:off_diag_H}
Z = \sum_{z} \sum_{ S_{ \mathbf{i}_q } = \identity } D_{( z, S_{ \mathbf{i}_q } )}e^{-\beta[ E_{z_0}, \ldots, E_{z_q}]}\,.
\end{equation}
The double sum above runs over all computational basis states $|z\rangle$ and all products $S_{\mathbf{i}_q}=P_{i_q} \ldots P_{i_2} P_{i_1}$ of permutation operators that evaluate to the identity. Here $q=0, \ldots, \infty$ denotes the number of elements in each product. Specifically, $ \mathbf{i}_q=(i_1, i_2, \ldots, i_q)$ is a $q$-element multi-index where each index $i_j$ ($j=1 \ldots q$) runs from $1$ to $M$. 

In the above sum, each summand is a product of two terms. The first is $D_{ ( z, S_{\mathbf{i}_q} ) } \equiv \prod_{j=1}^{q} d_{z_j}^{(i_j)}$ consisting of a product of the matrix elements
\begin{equation}
d_{z_j}^{(i_j)} = \langle z_j \vert D_{i_j} \vert z_j \rangle \,. 
\end{equation}
The various $\{|z_j\rangle\}$ states are the states obtained from the action of the ordered $P_j$ operators in the product $S_{{\bf{i}}_q}$ on $|z_0\rangle$, then on $|z_1\rangle$, and so forth. For example, for $S_{{\bf{i}}_q}=P_{i_q} \ldots P_{i_2}P_{i_1}$, we obtain $|z_0\rangle=|z\rangle, P_{i_1}|z_0\rangle=|z_1\rangle, P_{i_2}|z_1\rangle=|z_2\rangle$, etc. The proper indexing of the states $|z_j\rangle$ along the path is \hbox{$|z_{(i_1,i_2,\ldots,i_j)}\rangle$} to indicate that the state in the $j$-th step depends on all $P_{i_1}\ldots P_{i_j}$. For conciseness, we will use the shorthand $|z_j\rangle$. The sequence of basis states $\{|z_j\rangle \}$ may be viewed as a closed `walk' on the Hamiltonian graph -- the graph defined by $H$ such that the $H_{ij}$ matrix element corresponds to an edge between the two basis states $i$ and $j$, which serve as nodes on the graph. 

The second term in each summand, $e^{-\beta[ E_{z_0}, \ldots, E_{z_q}]}$,  is called the divided differences of the function \hbox{$F(\cdot) = e^{-\beta (\cdot)}$} with respect to the inputs $[ E_{z_0}, \ldots, E_{z_q}]$. The divided differences~\cite{Whittaker67,de05} of a function $F[\cdot]$ is defined as,
\begin{equation}
F[ E_{z_0}, \ldots, E_{z_q} ] \equiv \sum_{j=0}^{q} \frac{ F(E_{z_j}) }{ \prod_{k \neq j} ( E_{z_j} - E_{z_k} ) } \,.
\end{equation}
In our case, the inputs $E_{z_j}$ are defined as \hbox{$E_{z_j} = \langle z_j \vert D_0 \vert z_j \rangle$}. The reader is referred to Appendix~\ref{apx:odp} for additional details.

\subsection{PMR of the Bose-Hubbard model}

The Bose-Hubbard Hamiltonian, which is the focus of this study, is given by
\begin{equation}
H = -t \sum_{m=1}^M{ \hat{b}_{j_m}^{\dag}\hat{b}_{k_m} } 
+\frac{U}{2} \sum_{i=1}^L{ \hat{n}_i ( \hat{n}_i - 1 ) } - \mu \sum_{i=1}^L{ \hat{n}_i  } \,,
\end{equation}
where in the above expression $i=1, \ldots, L$ labels the sites, which we will treat as graph nodes for reasons that will become clear later, and $m=1, \ldots, M$ labels the (directed) `edges' of the model, i.e., the ordered pairs of sites $(j_m,k_m)$ between which hopping terms ${ \hat{b}_{j_m}^{\dag}\hat{b}_{k_m} } $ exist. In addition, hermiticity of the Hamiltonian dictates that for every  pair of indices $(j_m,k_m)$ there exists another pair $(j_{m'},k_{m'})$ such as $(j_{m'},k_{m'}) = (k_{m},j_{m})$, corresponding to a hopping term in the opposite direction. 

As the computational basis for the PMR expansion, we use the second quantized occupation number basis for bosons, where a basis state is given as $|{\bf n}\rangle = \vert n_1,n_2,\ldots,n_L \rangle$ with $L$ being the number of sites and $n_1, \ldots, n_L$ are nonnegative integers representing the number of bosons in each site. We denote the total number of bosons, $\sum_{i=1}^L n_i$, by $N$. 
The operators $\hat{b}_{i}^{\dag}, \hat{b}_{i}$ are creation and annihilation operators, respectively, obeying
\begin{equation}
\hat{b}_{i}^{\dag} \hat{b}_{j} |{\bf n}\rangle  = \sqrt{(n_i+1)n_j}  |{\bf n}^{(i,j)}\rangle \,,
\end{equation}
where $|{\bf n}^{(i,j)}\rangle$ stands for the state $|{\bf n}\rangle$ with one additional boson at site $i$ and one fewer at site $j$. The operator $\hat{n}_i=\hat{b}_{i}^{\dag} \hat{b}_{i}$ is the number operator. The coefficients $t,U$ and $\mu$ are real-valued parameters. 

Casting $H$ in PMR form with respect to the second quantized basis dictates that the diagonal term $D_0$ consists of the on-site terms, namely,
\begin{equation}
D_0 = \frac{U}{2}\sum_{i}{\hat{n}_i ( \hat{n}_i - 1 ) } - \mu \sum_i{ \hat{n}_i } \,.
\end{equation}
Likewise, the generalized permutation operators of the BH model are \hbox{$\tilde{P}_m=-t \hat{b}_{j_m}^{\dag}\hat{b}_{k_m}$}. These can be written as products of bonafide permutation operators which obey
\begin{equation}
P_m|{\bf n}\rangle  =  |{\bf n}^{(j_m,k_m)}\rangle \,,
\end{equation}
and accompanying diagonal operators
\begin{equation}
D_m = -t \sum_{{\bf n}} \sqrt{n_{j_m}(n_{k_m}+1)} |{\bf n}\rangle \langle {\bf n} |\,,
\end{equation}
which together give $\tilde{P}_m = D_m P_m$. Here, the summation index $\bf n$ runs over all basis states (though in the case where the number of particles is conserved, the sum of over states ${\bf n}$ can be restricted to those states that obey $\sum_{i=1}^L n_i = N$).
The total Hamiltonian can now be recast as
\begin{equation}
H =D_0  + \sum_{m=1}^M D_m P_m  \,.
\end{equation}
Using the above notation, the partition function can be written as 
\begin{equation}\label{eq:zbh}
Z= \sum_{{\bf n}} \sum_{{\bf{i}}_q} W_{({\bf n},S_{{\bf{i}}_q})}  = 
\sum_{{\bf n}} \sum_{{\bf{i}}_q}  D_{( \n, S_{ \mathbf{i}_q } )}e^{-\beta[ E_{\n_0}, \ldots, E_{\n_q}]}
\,.
\end{equation}
As already discussed, the operator sequences are of the form \hbox{$S_{{\bf{i}}_q}=P_{i_q} \ldots P_{i_2}P_{i_1}$} and must evaluate to the identity operation. 
Each $S_{{\bf{i}}_q}$ generates a sequence of states $|\n_0\rangle=|\n\rangle, P_{i_1}|\n_0\rangle=|\n_1\rangle, P_{i_2}|\n_1\rangle=|\n_2\rangle$ and so on where the last state is $|\n_q\rangle=|\n_0\rangle$. Moreover,  ${\displaystyle D_{(\n,S_{{\bf i}_q})} = \prod_{r=1}^{q} d_{\n_r}^{(i_r)} }$, 
where 
\begin{equation}
\label{eq:D}
	d_{\n_r}^{(m)}  = \langle \n_r|D_m|\n_r\rangle = -t\sqrt{n^{(r)}_{j_m}(n^{(r)}_{k_m}+1)} \,.
\end{equation}
Here, $n^{(r)}_{i}$ refers to the $i$-th element of the state $|\n_r\rangle$.

\subsection{The algorithm\label{section:qmc_algorithm}}

\subsubsection{Preliminaries}
Having presented the partition function as a sum of efficiently computable terms [Eq.~\eqref{eq:zbh}], we can now devise a QMC algorithm, i.e., a Markov chain Monte Carlo process, based on this decomposition. The partition function has the form of a sum configuration weights 
\begin{equation}
Z = \sum_{{\cal C}} W_{\cal C},
\end{equation}
where the weights are given by 
\begin{equation}
W_{\cal C} = D_{ (\n, S_{{\bf i}_q}) } e^{ - \beta [ E_{\n_0}, \ldots, E_{\n_q} ] }\,,
\end{equation}
and each configuration ${\cal C}$ is the pair ${\cal C}=\{ \vert \n \rangle, S_{\mathbf{i}_q} \}$. Here, $\vert \n \rangle$ is the basis state of the configuration and $S_{\mathbf{i}_q}$ is a product of operators that evaluates to $\identity$. As already discussed, each configuration ${\cal C}$ induces a closed walk on the Hamiltonian graph, a sequence of states 
\hbox{$\vert \n \rangle=\vert \n_0 \rangle, \vert \n_1 \rangle, \ldots, \vert \n_q \rangle=\vert \n \rangle$} which is acquired by acting with the permutation operators in $S_{\mathbf{i}_q}$, in sequence, on $\vert \n \rangle$. 

\subsubsection{The initial configuration}

The initial configuration of our QMC algorithm is set to be ${\cal C}_0 = \{ \vert \n \rangle, S_0 = \identity \}$, where $\vert \n \rangle$ is a randomly chosen basis state acquired by acting with a predetermined number of randomly picked operators $P_i$ on a predetermined basis state $\vert \n \rangle$, which we choose to be $\vert \n \rangle = \vert N, 0, 0, \ldots, 0 \rangle $, where $N$ is the total number of particles chosen for the initial state. The sequence of permutation operators is simply the empty sequence, for which $q=0$. 
The weight of the initial state is therefore given by $W_{{\cal C}_0} = e^{ -\beta [ E_{\n} ] } = e^{ -\beta E_{\n} }$.

\subsubsection{The QMC updates\label{section:insertion_deletion}}
To ensure that every configuration in configuration space is reachable from any other, i.e., that the Markov chain is ergodic, we utilize five different types of moves. These are (i) `classical' moves, (ii) local swap moves (iii) cyclic rotation moves, (iv) block swaps and (v) insertion-deletion moves. We discuss these in detail below. We then show that this set of moves together is sufficient to guarantee ergodicity. 

\noindent {\bf Classical moves.}--- 
Classical moves ensure that all basis states $|\n\rangle$ can be reached. During this move, a new basis state $\vert \n^\prime \rangle$ is proposed to replace the current one $\vert \n \rangle$ in the configuration ${\cal C}$.  The sequence of operators $S_{{\bf i}_q}$ is not altered. {Our algorithm may work both in the canonical ensemble and in the grand-canonical ensemble. In a canonical ensemble treatment, updates may be easily adjusted so as to conserve the number of bosons in the system. Otherwise classical updates may change the total number of particles. If working within a specific particle number sector (i.e., in the canonical ensemble), the new proposed basis state may be chosen such the total number of bosons is conserved. This can be achieved \ by acting with a randomly selected permutation operator $P_m$ on the current basis state. A non particle-number-conserving  move may consist of adding or removing a boson from a randomly chosen lattice site.}
In the case where the proposed new state $\vert \n^\prime \rangle$ is not a valid state, i.e., whenever $P_m|\n\rangle=0$, the procedure is repeated until a valid state is produced. The new configuration is accepted with probability $\min(1,W_{{\cal C}'}/W_{\cal C})$ where $W_{{\cal C}'}$ is the weight of the proposed configuration ${\cal C}'$ and $W_{\cal C}$ is the weight of the current one ${\cal C}$. 

\noindent {\bf Local swap moves.}--- 
A local swap move consists of randomly picking two adjacent operators in $S_{{\bf i}_q}$ and then swapping them to create a new sequence $S'_{{\bf i}_q}$. Here too, the new configuration is accepted with probability $\min(1,W_{{\cal C}'}/W_{\cal C})$ where $W_{{\cal C}'}$ is the weight of the proposed configuration ${\cal C}'$ and $W_{\cal C}$ is the weight of the current one ${\cal C}$. 

\noindent {\bf Cyclic rotation moves.}--- 
The cyclic rotation move consists of rotating (typically small length) sub-sequences within $S_{{\bf i}_q}$ that evaluate to $\identity$ -- we shall refer to these as cycles -- utilizing the fact that a rotated sub-sequence that evaluates to $\identity$ also evaluates to $\identity$. The chosen sub-sequence $S$ is virtually `cut' to two so that it can be written as $S=S_1 S_2$. Then, $S$ is replaced with the modified sub-sequence $S'=S_2 S_1$ in  $S_{{\bf i}_q}$. Here too, the new configuration is accepted with probability $\min(1,W_{{\cal C}'}/W_{\cal C})$ where $W_{{\cal C}'}$ is the weight of the proposed configuration ${\cal C}'$ and $W_{\cal C}$ is the weight of the current one ${\cal C}$.

\noindent {\bf Block swap moves.}--- The block swap move modifies both the basis state and the sequence of operators. 
Here, a random position $k$ in the product $S_{{\bf{i}}_q}$ is picked such that the product is split into two (non-empty) sub-sequences, $S_{{\bf{i}}_q}=S_2 S_1$, with $S_1 = P_{i_k} \cdots P_{i_{1}}$ and $S_2 = P_{i_{q}} \cdots P_{i_{k+1}}$.  Denoting the classical state  at position $k$  in the product as $\vert {\bf n}' \rangle$, namely,
\begin{equation}
|{\bf n}'\rangle=S_1|z\rangle=P_{i_k} \cdots P_{i_1}|{\bf n}\rangle \,,
\end{equation}
where $\vert \n \rangle$ is the classical state of the current configuration, the new block-swapped configuration is \hbox{${\cal C}'=\{|\n'\rangle, S_1 S_2\}$}.

\noindent {\bf Insertion-deletion moves.}---
The insertion-deletion move is the only type of move considered here that changes the length $q$ of the sequence of operators. An insertion-deletion move either removes cycles (sequences of operators that evaluates to the $\identity$)  from $S_{{\bf{i}}_q}$ or inserts a randomly picked cycle from a pool of `fundamental cycles' (which we discuss in detail in the next section).

The insertion-deletion move consists of first randomly selecting a length $m_l$ for the cycle that is to be inserted or removed among all possible cycle lengths. As the next step, a random choice is made as to whether insert a cycle or remove one from $S_{{\bf{i}}_q}$ .

If deletion is selected, and $m_l = 2$, a uniformly random deletion point $k$ is selected. If $P_{i_{k-1}} P_{i_k}$ is a cycle, i.e., evaluates to the identity operation, then a configuration with the two operators removed is proposed. Otherwise, the move is rejected. For $m_l > 2$, a deletion point $k$ is selected in a similar manner. If $\{P_{i_{k-2}},P_{i_{k-1}}, \cdots, P_{i_{k+m_l-3}}\}$ is equivalent to $\identity$ and the sequence is in the list of fundamental cycles, the subsequence is removed and the resultant configuration is proposed. Otherwise, no new configuration is proposed and the move is rejected.

If insertion is selected, a random insertion point $k$ is selected. A random cycle of length $m_l$ is picked from the pool of cycles which is then inserted into the full sequence $S_{{\bf{i}}_q}$ at position $k$. 
The proposed new configuration is then accepted or rejected based on its relative weight (and other selection factors) maintaining detailed balance.

\noindent {\bf Cycle completion.}---
Although not strictly necessary for ergodicity, one may augment the aforementioned QMC updates with another type of moves, which we refer to here as `cycle completion moves'. 
Here, one chooses a sub-sequence $S_1$ from $S_{{\bf{i}}_q}$ and subsequently checks whether $S_1$ is a sub-cycle of one the aforementioned fundamental cycles, namely if a fundamental cycle of the form $S_1 S_2 = \identity$ exists. If it does, then $S_1$ is replaced (with the appropriate acceptance probability) with $S_2^{-1}$ as both $S_1$ and its replacement evaluate to the same permutation. 

At this point, it would be worthwhile to contrast the updates of the present algorithm against those usually used in existing techniques. It is interesting to note that while existing approaches such as Stochastic Series Expansion (SSE)~\cite{Kawaki17,doi:10.1143/JPSJS.74S.10}  or continuous-time path integral Monte Carlo-based methods~\cite{Pollet_2012} require world line-type or worm-type updates where `disturbances' along the imaginary time dimension are created and then healed in order to create new configurations, PMR-QMC does not require such updates. This is for two main reasons. The first of which is that unlike existing schemes the PMR-QMC quantum imaginary-time dimension consists only of off-diagonal operators (permutation operators) as the diagonal component of the Hamiltonian is explicitly integrated out (diagonal matrix elements appear only as divided-difference coefficients). Second, the insertion-deletion of either pairs of operators or fundamental cycles along the sequence of operators, i.e., along the imaginary time dimension function as a short-distance worm thereby minimizing the risk of percolation. 

Nonetheless, worm-type moves in the framework of PMR-QMC may also be implemented although as mentioned, they are not strictly necessary. A worm update would introduce a ‘disturbance’
(or a ‘worm head’) into the sequence of operators $S_{{\bf{i}}_q}$ by either inserting into $S_{{\bf{i}}_q}$ a single permutation operator or removing one from it (we will call this addition or removal of an operator a ‘single operator move’). An insertion or removal of a single permutation operator causes the disturbed sequence to evaluate to a non-identity permutation and hence corresponds to a zero-weight configuration. As a result, the disturbed sequence must be ‘healed’ in order to form a sequence that evaluates to the identity. The healing process proceeds by introducing additional moves: either employing standard local updates such as the ones already discussed (namely, local swap, cycle completion, and cycle rotation) or additional single operator moves. These single operator moves have the power to heal the sequence. After every such move, the instantaneous sequence can be checked to determine whether it evaluates to the identity operator. If it does, the worm update can be terminated. If it does not, additional moves are required. To make sure that detailed balance is conserved and that eventual acceptance rates of the intermediate worm moves are high, we assign non-identity intermediate configurations their ‘natural’ weight $W_{\cal C}$. 

\section{Measurements\label{sec:measurements}}

Deriving expressions for measurements of expectation values of essentially any physical observable is straightforward with PMR~\cite{advancedMeasurements}. Below we provide a number of useful examples, including various energy measurements, arbitrary functions of the Hamiltonian and local observables. In addition, we discuss the measurement of quantities that are of particular importance to the Bose-Hubbard model such as superfluid density and the one-body density matrix. 

For all of above, the basic idea would be to write the thermal average of any given operator $A$ as
\beq
\langle A\rangle = \frac{\tr[A\e^{-\beta H}]}{\tr\e^{-\beta H}]}=\frac{\sum_{{\cal C}} A_{{\cal C}} W_{{\cal C}} }{\sum_{{\cal C}}W_{{\cal C}} } \,.
\eeq
The quantity $A_{{\cal C}}$ is therefore the instantaneous quantity associated with the configuration  ${{\cal C}}$ that should be calculated and stored during the simulation. Since the configurations are visited in proportion to their weights, a simple average of the above quantities will yield the correct expectation values for the diagonal, off-diagonal and total energies respectively. 

\subsection{Energies}
The average energy $\langle H \rangle$ may be calculated using the expression:
\begin{equation}
\label{eq:H}
\langle H \rangle = \frac{\tr[H e^{-\beta H}]}{\tr[e^{-\beta H}]}=\frac{ \sum_{{\cal C}} W_{(\n,S_{{{\bf i}_q}})}\left(E_\n + \frac{e^{-\beta[E_{\n_1},\ldots,E_{\n_q}]}}{e^{-\beta[E_\n,\ldots,E_{\n_q}]} }\right)}{ \sum_{{\cal C}} W_{(\n,S_{{{\bf i}_q}})}}\,. 
\end{equation}
In the above expression we identify $E_\n$ as the instantaneous quantity that needs to be calculated for the diagonal component of the Hamiltonian throughout the simulation, namely, 
\begin{equation}
\label{eq:H_diag}
\langle H_{\textrm{diag}} \rangle = \frac{\tr[H_{\textrm{diag}} e^{-\beta H}]}{\tr[e^{-\beta H}]}=\frac{ \sum_{{\cal C}} W_{(\n,S_{{{\bf i}_q}})}\ E_\n}{ \sum_{{\cal C}} W_{(\n,S_{{{\bf i}_q}})}}\ \,,
\end{equation}
and $\frac{e^{-\beta[E_{\n_1},\ldots,E_{\n_q}]}}{e^{-\beta[E_\n,\ldots,E_{\n_q}]} }$ as the quantity corresponding to the off-diagonal component of the Hamiltonian, that is: 
\begin{equation}
\label{eq:H_od}
\langle H_{\textrm{off-diag}} \rangle = \frac{\tr[H_{\textrm{off-diag}} e^{-\beta H}]}{\tr[e^{-\beta H}]}=\frac{ \sum_{{\cal C}} W_{(\n,S_{{{\bf i}_q}})} \frac{e^{-\beta[E_{\n_1},\ldots,E_{\n_q}]}}{e^{-\beta[E_\n,\ldots,E_{\n_q}]} }}{ \sum_{{\cal C}} W_{(\n,S_{{{\bf i}_q}})}}.
\end{equation}
The sum of these two instantaneous quantities yields the instantaneous total energy. 

\subsection{General functions of the Hamiltonian}

Expectation values for more general functions of the Hamiltonian, namely,
\beq
\langle g(H) \rangle = \frac{\tr[g(H) \e^{-\beta H}]}{\tr[ \e^{-\beta H}]} \,,
\eeq
may be obtain by applying the off-diagonal series expansion to $\tr[g(H) \e^{-\beta H}]$ which yields~\cite{advancedMeasurements}:
\beq
\langle g(H) \rangle =\frac
{
\sum_{{\cal C}}  W_{(\n,S_{{{\bf i}_q}})} 
\left( \sum_{j=0}^q g[E_{\n_0},\ldots,E_{\n_j}] 
\frac{\e^{-\beta[E_{\n_j},\ldots,E_{\n_q}]}}{\e^{-\beta[E_{\n_0},\ldots,E_{\n_q}]}} \right)
}
{ \sum_{{\cal C}} W_{(\n,S_{{{\bf i}_q}})}} \,,
\eeq
where $g[E_{\n_0},\ldots,E_{\n_j}]$ is the divided difference with respect to the function $g(\cdot)$. 
Given the above expression, we may identify \hbox{$\sum_{j=0}^q g[E_{\n_0},\ldots,E_{\n_j}] 
\frac{\e^{-\beta[E_{\n_j},\ldots,E_{\n_q}]}}{\e^{-\beta[E_{\n_0},\ldots,E_{\n_q}]}}$} as the quantity that is to be evaluated and collected during the QMC simulation. 

In the special case where powers of the Hamiltonian, $H^k$, are considered, we get
\beq
\langle H^k \rangle =\frac
{
\sum_{{\cal C}} W_{(\n,S_{{{\bf i}_q}})} 
\left( \sum_{j=0}^{\max\{k,q\}} [E_{\n_0},\ldots,E_{\n_j}]^k 
\frac{\e^{-\beta[E_{\n_j},\ldots,E_{\n_q}]}}{\e^{-\beta[E_{\n_0},\ldots,E_{\n_q}]}} \right)
}
{ \sum_{{\cal C}} W_{(\n,S_{{{\bf i}_q}})}}\,,
\eeq
which follows from the fact that $[E_{\n_0},\ldots,E_{\n_j}]^k $ evaluates to $0$ for $j>k$, for $j=k$ it evaluates to $1$, and in general for $k \geq j$ and for arbitrary inputs $x_0,\ldots,x_j$:
\beq
[x_0,\ldots,x_j]^k  = \sum_{a\in\{0,\dots,n\}^{k-j} \text{ with } a_1 \le a_2 \le \dots \le a_{k-j}} \prod_{m\in a} x_m.
\eeq

\subsection{Measurements of arbitrary static operators}
\label{subsec:measurements-of-arb-static-op}

We next consider the measurement of a general static operator $A$. We proceed by casting it in PMR form, i.e., as $A=\sum_i \tilde A_i \tilde P_i$ where each $\tilde A_i$ is diagonal and the $\tilde P_i$'s are either permutation operators that appear in the Hamiltonian or products thereof. In this case, we can write
\beq \label{eq:AiPi}
\langle A \rangle = \frac{\tr[A \e^{-\beta H}]}{\tr[\e^{-\beta H}]}= \sum_i \frac{\tr[\tilde A_i \tilde P_i \e^{-\beta H}]}{\tr[\e^{-\beta H}]} \,,
\eeq
and we may therefore focus on a single $\tilde A \tilde P$ term at a time. 
Carrying out the off-diagonal expansion, we end up with:
 \bea \label{eq:gen1}
&&\tr[\tilde{A} \tilde{P} \e^{-\beta H}]= \sum_{\n} \tilde{A}(\n)\sum_{q=0}^{\infty}  \sum_{{S_{{\bf{i}}_q}}}  D_{(\n,S_{{{\bf i}_q}})}  \nonumber\\
&\times& {\e^{-\beta [E_{\n_0},\ldots,E_{\n_q}]}} \langle \n| \tilde{P} S_{{{\bf i}_q}} \ket{\n} \,. 
\eea
where $D_{(\n,{S}_{{{\bf i}_q}})}  {\e^{-\beta [E_{\n_0},\ldots,E_{\n_q}]}}$ is the weight of the configuration $\{\n,{S}_{{{\bf i}_q}}\}$.

The operator to be measured has the form $A=\tilde{A} \tilde{P}$ where $\tilde{A}$ is diagonal and 
$\tilde{P} = P_{i_1} P_{i_2} \cdots P_{i_k}$.
We modify Eq.~(\ref{eq:gen1}) so that $(\n,\tilde{S}_{{{\bf i}_q}})$ with  $\tilde{S}_{{{\bf i}_q}} = \tilde{P} S_{{{\bf i}_q}}$ is seen as a configuration instead of $(\n,{S}_{{{\bf i}_q}})$.
Thus, we arrive at: 
\beq
\langle A \rangle = 
\frac
{
\sum_{(\n,{\tilde{S}_{{\bf{i}}_q}})} w_{(\n,{{\tilde{S}_{{\bf{i}}_q}}})} 
 M_{\tilde A \tilde P}{(\n,{\tilde{S}_{{\bf{i}}_q}})}  
}
{ \sum_{(\n,{\tilde{S}_{{\bf{i}}_q}})} w_{(\n,{{\tilde{S}_{{\bf{i}}_q}}})}},
\label{eq:customOperator1}
\eeq
where
\beq
M_{\tilde{A}\tilde{P}}{(\n,{\tilde{S}_{{\bf{i}}_q}})} = \delta_{\tilde{P}} \tilde{A}(\n) \frac{1}{D_{(\n,\tilde{P})}} 
\frac{  {\e^{-\beta [E_{\n_0},\ldots,E_{\n_{q-k}}]}} } { {\e^{-\beta [E_{\n_0},\ldots,E_{\n_{q}}]}} } \,.
\label{eq:customOperator2}
\eeq
In the above, $\delta_{\tilde{P}}=1$ if the leftmost operators of $\tilde{S}_{{{\bf i}_q}}$ are $P_{i_1} P_{i_2} \cdots P_{i_k}$ and is zero otherwise,
and 
\beq
D_{(\n,\tilde{P})} = \frac{D_{(\n,\tilde{S}_{{{\bf i}_q}})}}{D_{(\n,{S}_{{{\bf i}_q}})}} = \prod_{m=1}^{k} \langle \n_{q-m+1} |D_{i_m}| \n_{q-m+1} \rangle.
\eeq

One important example of operators of the above form are the matrix elements of the 
so-called one-body density matrix, which establishes a condensation criterion in terms of the  properties of the matrix whose elements are
$\rho_{ij} = \langle b_i^\dagger b_j \rangle$~\cite{OBDM,penrose1,penrose2}. 
{Operators of the form $b_i^\dagger b_j$ can be written as products of the form
\hbox{$(b_i^\dagger b_k) (b_k^\dagger b_m) \cdots (b^\dagger_l b_j)$} or permutations thereof, where each of the operators \hbox{$b_i^\dagger b_k, b_k^\dagger b_m, \ldots, b^\dagger_l b_j$} corresponds to a permutation operator appearing in the Hamiltonian and can therefore be readily measured using PMR-QMC. We note that for any given matrix element $\rho_{ij}$ there will be multiple distinct `paths', or products of operators, that  evaluate to the target operator $b_i^\dagger b_j$ all of which can be taken to contribute to the statistics of $\rho_{ij}$.  In cases where the graph distance between site $i$ and site $j$ is long, any particular product may have low likelihood to be encountered, however in this case there will be in general factorially many paths between the two sites, all of which can be taken into account. }

\subsection{Calculating the superfluid density}

The concept of superfluid density~\cite{Leggett} in the Bose-Hubbard model is particularly important when studying phase transitions in ultracold atomic systems. It provides insight into the coherent motion of particles and is a key quantity in characterizing the different quantum phases of the system. 
In cases where the number of particles is conserved, in which case the winding number is well defined,
measurement of the superfluid density can be directly connected to fluctuations of the winding number~\cite{pollock1987path}. In the most general case, the superfluid density is proportional to
\beq
\rho_s \propto \left\langle \left( \sum_j L_j W_j \hat{r}_j \right)^2 \right\rangle \,,
\eeq
where $L_j$ is the linear size of the system is the $\hat{r}_j$ direction and the quantity $W_j$ counts the number of particles that cross the boundaries of the system in direction $\hat{r}_j$. 
For instance, in the special case of a $d$-dimensional hypercubic lattice, with $L^d$ sites, 
one obtains
\beq
\rho_s = \frac{L^{2-d}}{2 \beta t d} \left\langle \sum_{j=1}^d W_j^2 \right\rangle \,,
\eeq
and where more complex geometries require the calculation of other bilinear combinations of $W_j$. 
Collecting statistics for $W_j$, number of particles that cross the boundaries of the system in the $j$-th direction, in PMR-QMC is simple. Since every operator in the sequence of operators $S_{{{\bf i}_q}}$ corresponds to a directed edge, $W_j$ is given by $W_j=N_j^+ - N_j^-$ where $N_j^+$ counts the number of edges that cross the boundary in the positive $j$ direction and $N_j^-$ counts the number of edges that cross the boundary in the negative $j$ direction. 

\section{Ergodicity and minimal cycle bases\label{sec:mcb}}

The QMC update moves used throughout the simulation must be able to generate an ergodic Markov chain for any input graph and dimensionality of the BH model. That is, any valid configuration $(|\n\rangle, S_{{\bf i}_q})$ has to be reachable from any other.  While the various (second-quantized) basis states $|\n\rangle$ are trivially reachable from one another by the so-called `classical moves' discussed in the previous section, which randomly alter the basis states (augmented by block swap moves, which also change the basis state), less obvious is the guarantee that all operator sequences  $S_{{\bf i}_q}$ evaluating to the identity  are reachable from one another.

To show that the moves discussed in the previous section do indeed generate an ergodic Markov chain, we begin by making a few observations. The first is that local swap and cyclic rotation moves shuffle, or permute, the operators in the sequence of operators. Thus, to demonstrate ergodicity one only needs to show that all valid multi-sets of operators (irrespective of their ordering) are produceable.

The second observation we make is that every permutation operator $P_m$ in the BH model, which as already mentioned can be associated with a directed edge on the BH graph, has an inverse permutation $P_{m'}$ such that \hbox{$P_{m'}=P_m^{-1}$} -- the permutation operator associated with the same edge but which points in the opposite direction. 
The insertion-deletion move consisting of the insertion or deletion of pairs of operators $P_m P^{-1}_m$ therefore corresponds to the insertion and deletion of operators corresponding to the same edge (but with opposite directions) twice. The insertion-deletion of pairs can therefore be used to remove edge pairs down to a core collection of operators that multiply to the identity and in which operators do not appear with their inverses.
We conclude then, that to guarantee ergodicity, the only remaining requirement is that there is an update move capable of generating all multi-sets of operators (whose product evaluates to the identity) which contain edges pointing only in one direction but never both (that is, sequences that never contain both $P_m$ and $P_m^{-1}$). We shall call such multi-sets of operators `multi-cycles'. We shall call a multi-cycle  that does not contain repeated edges a `cycle' and note that any multi-cycle is a concentration of bonafide cycles. 

\begin{figure}[htp]
	\includegraphics[scale=0.5,trim={2.5cm 1.5cm 1.5cm 1.5cm},clip]{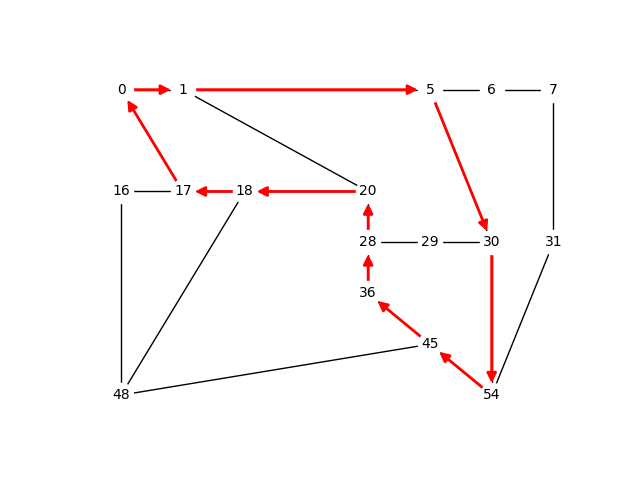}
	\caption{An example of a random graph on which the BH model can be defined. Nodes correspond to sites that the bosons can occupy and every edge is associated with two permutation operators, or hopping terms -- one in each direction. In red is an example of a set of (directed) edges whose corresponding sequence of operators multiply to the identity operation.}
	\label{fig:graph}
\end{figure}

In terms of edges on the BH graph, the ability to produce all multi-cycles reduces to the requirement that all cycles on the underlying BH graph can be produced, or inserted. An illustrative example of a single cycle on a BH graph is given in Fig.~\ref{fig:graph}. 

In what follows, we show that any cycle on a given BH graph can be produced via combinations of insertions and deletions of cycles taken from a finite set of cycles, commonly referred to as a cycle basis -- a set of cycles that combinations thereof are capable of producing all possible cycles~\cite{Berger04}. Setting up a QMC update rule within which these `fundamental'  cycles are inserted or deleted (see Sec.~\ref{section:insertion_deletion}) will ensure then that all cycles are produceable, guaranteeing ergodicity as desired. We next discuss the process of generating a cycle basis for any given input graph.

Let us consider a $K$-edge BH graph with $n$ sites labeled $1,\ldots,n$. The $M=2K$ permutation operators of the BH graph correspond to the directed edges, equivalently ordered pairs of nodes of the form $(j_m,k_m)$, corresponding to the existence of a permutation operator $P_m$ in the Hamiltonian which creates a boson at site $j_m$ and annihilates one at site $k_m$. A cycle $c$ (of length $|c|$) is a set of edges that can be ordered as a sequence 
$\{ (i_1,i_2), (i_2, i_3), \ldots, (i_{|c|},i_1)\}$ where $|c|$ denotes the number of edges in $c$, with the restriction that if an edge is in $c$ then its inverse cannot be in $c$. Succinctly, a cycle may be written as a sequence of  nodes $i_1 \to i_2 \to \cdots \to i_{|c|} \to i_1$.  

With the above definitions, one can assign every permutation operator $P_m$ corresponding to a directed edge $(j_m,k_m)$ a ternary vector ${\mathbf b }_m= (b_1, b_2, \ldots, b_n)$  such that $b_{j_m}=1$ (a boson is created at site $j_m$),  $b_{k_m}=-1$ (a boson is annihilated at site $k_m$) and all other entires are set to zero. 
The product of two permutation operators would correspond to the addition of two such vectors. A cycle $c$ would be a linear combination of ternary vectors adding up to the zero vector, namely, $\sum_{i=1}^M c_i {\bf b}_i = {\bf 0}$ where $c_i \in \{-1, 0, 1\}$.  

\begin{figure}[htp]
	\includegraphics[scale=0.2]{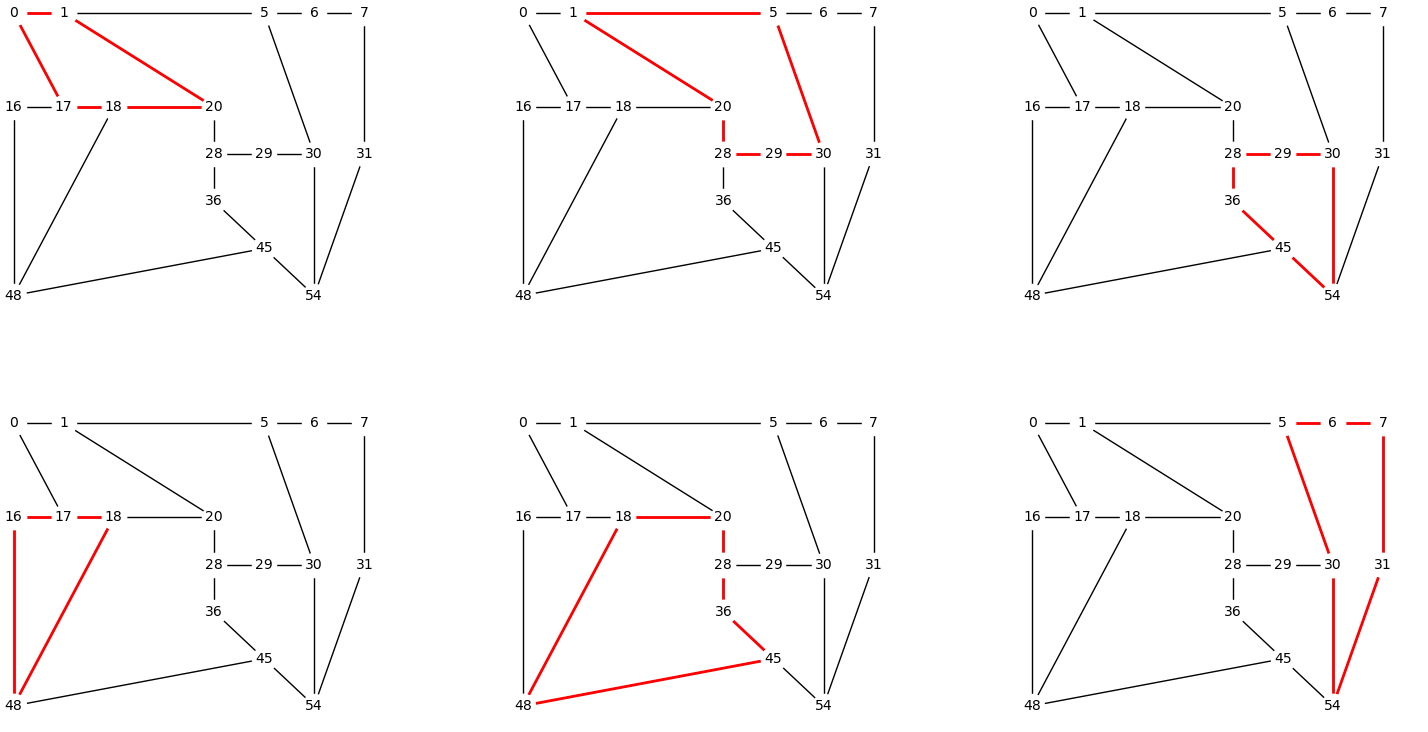}
	\caption{A cycle basis for the graph depicted in Fig.~\ref{fig:graph}. Every cycle on the BH graph can be represented as a combination (or a concatenation) of these basis cycles.}
	\label{fig:min_cycle_basis2}
\end{figure}

Finding a basis of cycles with which one could produce any possible cycle corresponds to finding a set of ternary vectors of the form  ${\bf c} = \{c_1, \ldots, c_M\}$ that solve the homogenous set of equations \hbox{${\bf B} {\bf c}= {\bf 0}$} where ${\bf B}$ is the $M \times n$ matrix consisting of the $M$ column vectors ${\bf b}_i$ ($i=1,\ldots, M$). Expressed differently, finding a cycle basis can be accomplished via finding the nullspace of the above linear system, which can be done efficiently using Gaussian elimination. 
In Fig.~\ref{fig:min_cycle_basis2}, we provide an example of a cycle basis found for the graph depicted in Fig.~\ref{fig:graph}. 
In the figure, a non-directed cycle is depicted as a collection of red-colored edges. 

Denoting by $T$ the dimension of the cycle nullspace, we note that the set of nullspace cycles is not unique, as any $T$ linearly independent vectors may serve as a basis. For the QMC algorithm however, we find that in order to maximize the acceptance ratios of insertion and removal of cycles the length of cycles should preferably be as short as possible. We therefore devise a protocol for producing a minimal cycle basis~\cite{depina95,Berger04,Kavitha08} -- the set of shortest possible cycles that form a basis.  We find the minimal cycle basis using an algorithm proposed by Kavitha et al.~\cite{Kavitha08}.
\begin{figure*}[htp]
\includegraphics[scale=0.28]{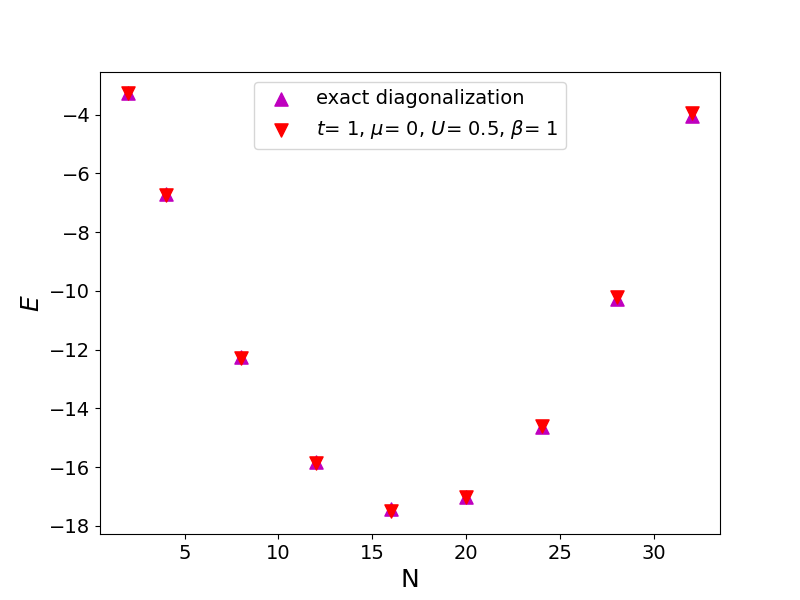}
\includegraphics[scale=0.28]{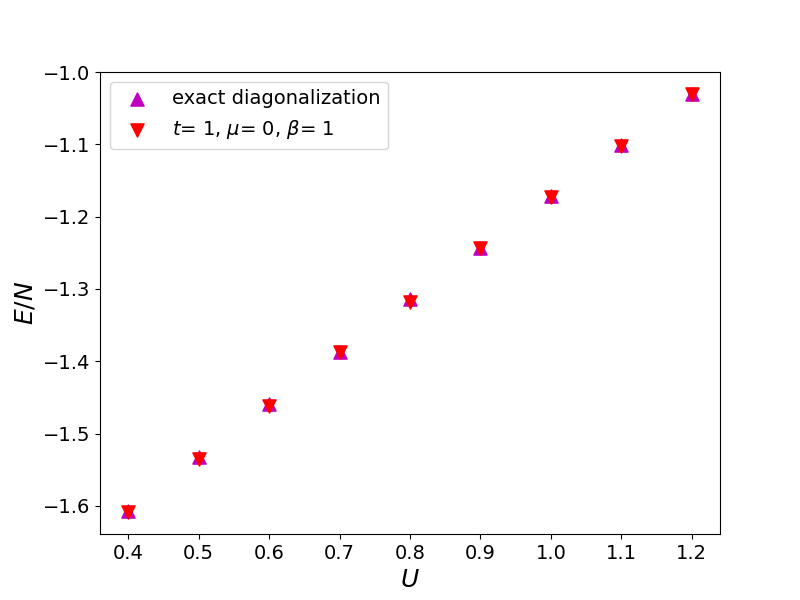}
\includegraphics[scale=0.28]{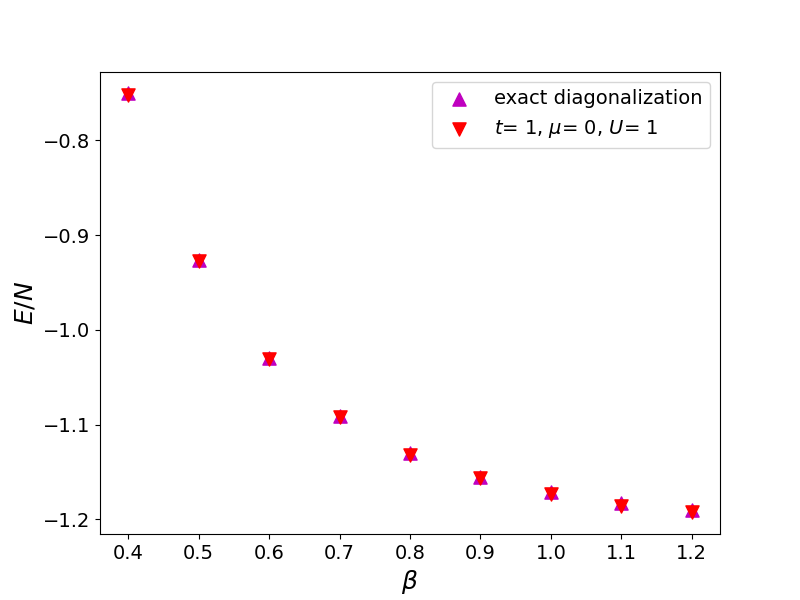}
\caption{Comparison of QMC results with exact diagonalization. Left: Average energy $E=\langle H \rangle$ as a function of total number of particles $N$ for a 2 by 2 rectangular lattice with open  boundary conditions and parameters $t=1, \mu=0, U=0.5, \beta=1$. Middle: Average energy $\langle H \rangle$ for a 2 by 2 rectangular lattice with $N=8$ particles (open boundary conditions) and parameters $t=1, \mu=0, \beta=1$ as a function of $U$. Right: Average energy for a 2 by 2 rectangular lattice with  $N=8$ particles (open boundary conditions) and parameters $t=1, \mu=0, U=1$ as a function of inverse temperature $\beta$.}
\label{fig:2x2_N}
\end{figure*}

We note that even though QMC updates based on the generation of a minimal cycle basis are sufficient to ensure an ergodic Markov chain, one may introduce additional cycles into the pool of `fundamental' cycles to improve the convergence rate of the simulation. Having more cycles in the pool of cycles available to choose from will increase the acceptance rates of both the insertion-deletion and cycle completion updates. On the other hand, searching a long list of fundamental cycles stands to inevitably slow down the algorithm. 
We find that these two opposing considerations are appropriately balanced if one includes all the chordless cycles of the BH graph that have a length smaller than or equal to the longest basis cycle found (a chordless cycle is defined as a cycle that does not have a `chord', i.e., a cycle for which there are no edges not belonging to the cycle that connect two vertices that do belong to it~\cite{Uno14}). 

\section{Algorithm testing\label{sec:results}}

To test the power and flexibility of our method, we have carried out QMC simulations for a variety of BH models, implementing the algorithm introduced above allowing it to find within each setup a minimal cycle basis and in turn provably ergodic QMC updates. We next present the results of our simulations for several BH graph configurations including rectangular lattices with varying Hamiltonian parameters as well as irregular graphs. For what follows, we have chosen to present the performance of the algorithm in the canonical ensemble. We have set the chemical potential $\mu$ to zero and have employed classical update moves that conserve the number of particles. 

\subsection{Verification against exact diagonalization}

To verify the correctness of our algorithm, we first carry out simulations of the BH model on small two-dimensional rectangular lattices so that the QMC results can be compared against those obtained from exact diagonalization.

For concreteness, we choose to monitor and measure the total energy, given in Eq.~(\ref{eq:H}). It should be noted that our algorithm is readily capable of measuring many other physical observables as well~\cite{advancedMeasurements}. All data points presented in this section were obtained via the execution of multiple independent simulations each of which yielding a single value for the total energy. Data points were obtained by averaging the values from each run whereas error bars were obtained by the evaluation of the sample error of the mean over said data points.

In Fig.~\ref{fig:2x2_N}(left), we plot the average thermal energy as a function of number of bosons $N$ for a BH model on a $2 \times 2$ rectangular lattice (with open  boundary conditions). The parameters for which results are shown are $t = 1, \mu = 0, U  = 0.5$ and $\beta = 1$. Figure~\ref{fig:2x2_N}(middle) shows the average energy as a function of the on-site repulsion $U$ for $N=8$ bosons. Here, $t = 1, \mu = 0$ and $\beta = 1$. Another set of results for simulations of a $2 \times 2$ rectangular lattice with open  boundary conditions is presented in Fig.~\ref{fig:2x2_N}(right). Here too, $N=8$ and the average thermal energy is plotted as a function of inverse-temperature $\beta$ (with $t = 1, \mu = 0$ and $U = 1$). As can be seen from the three panels of the figure, the QMC results are in excellent agreement with those obtained from exact diagonalization.

\subsection{Larger two-dimensional lattices}

\begin{figure}[htp]
\includegraphics[scale=0.42]{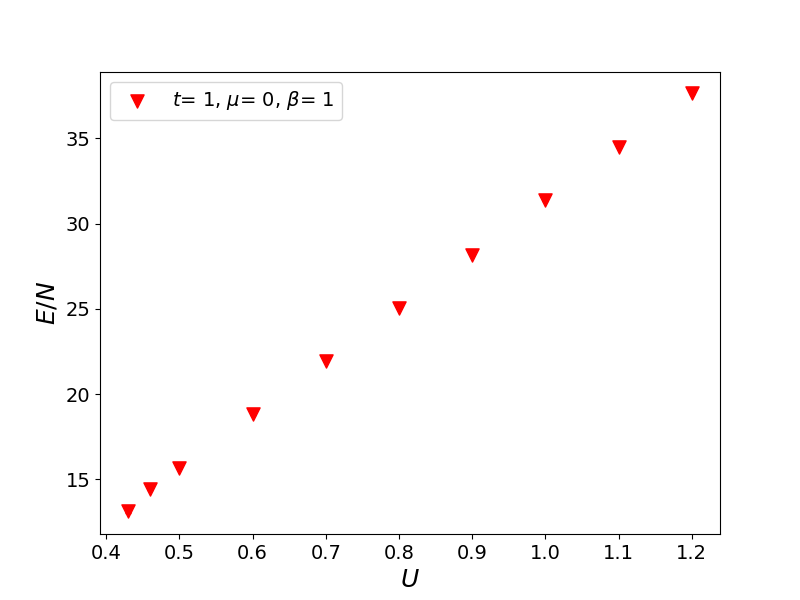}
\includegraphics[scale=0.42]{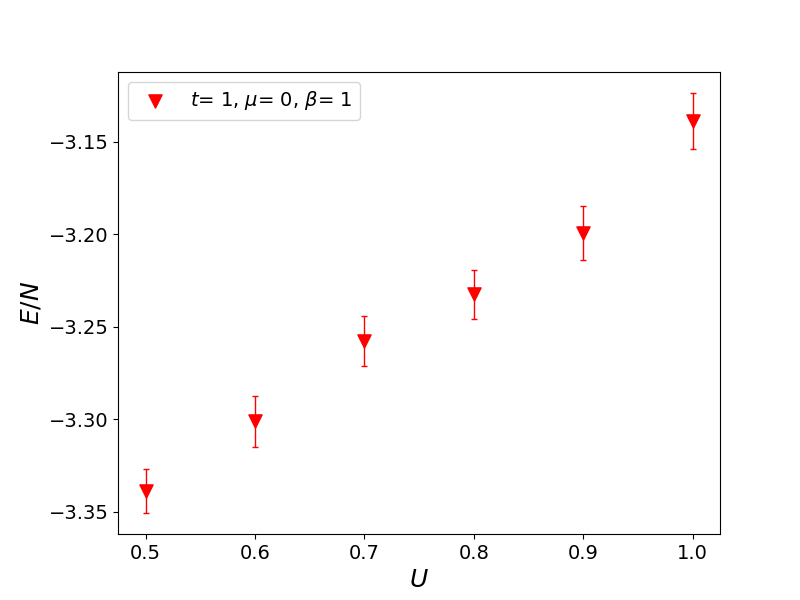}
\caption{Top: Average energy $E=\langle H \rangle$ for a BH model defined on an $8 \times 8$ rectangular lattice  with open  boundary conditions and $N=64$ particles as a function of on-site potential $U$. Here, $t=1, \mu=0$ and $\beta=1$. Bottom: Average energy $E=\langle H \rangle$ as a function of $U$ for a $6 \times 6$ rectangular lattice with periodic boundary conditions and 36 particles. Here too, $t=1, \mu=0$ and $\beta=1$.}
\label{fig:6x6_pbc_U} 
\end{figure}

Having verified the validity of our approach, we next provide simulation results for larger rectangular systems.
Figure~\ref{fig:6x6_pbc_U}(top) depicts the average thermal energy as a function of the on-site repulsion $U$ for a BH model defined on an $8 \times 8$ rectangular lattice with open boundary conditions containing $N=64$ particles. 
The average thermal energy is plotted as a function of  of on-site potential $U$ for a $6 \times 6$ rectangular lattice with periodic boundary conditions in Fig.~\ref{fig:6x6_pbc_U}(bottom). Here, $t = 1, \mu = 0$ and $\beta = 1$. 

\subsection{Simulations of the BH model on random graphs}

To showcase the versatility of our approach we have also carried out QMC simulations of BH models defined on randomly generated graphs.
For the results below, we present the graphs themselves and their fundamental basis cycles alongside the simulation results. 

\begin{figure*}[htp]
\includegraphics[scale=0.75]{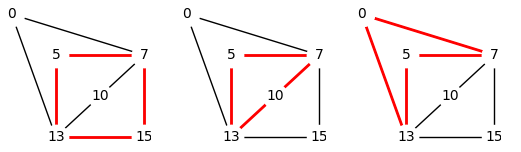}
\includegraphics[scale=0.3]{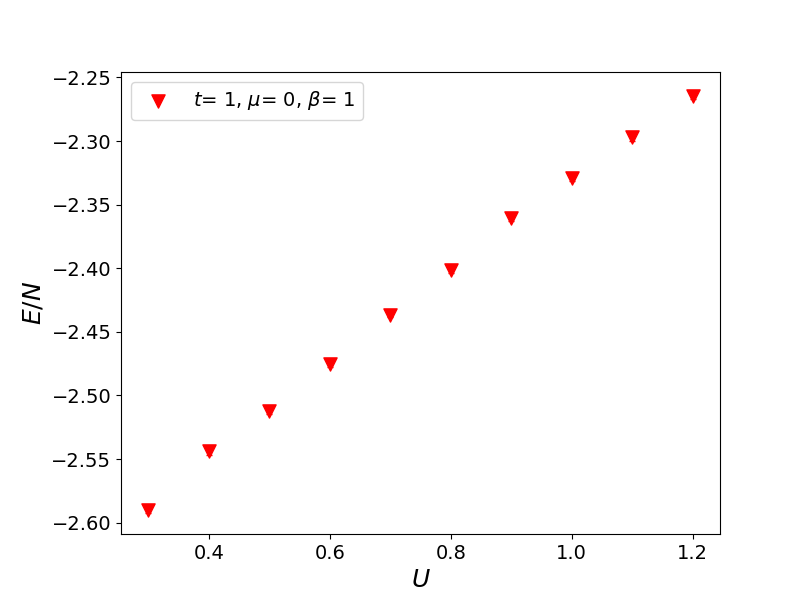}
\caption{Left: The minimal cycle basis (in red) for a six-node graph containing eight edges. Right: Average energy $E=\langle H \rangle$ as a function of $U$ for the graph depicted in the left panel. Here, the number of particles is $N=6$. The remaining parameters are fixed and have the following values: $t=1, \mu=0$ and $\beta=1$.}
\label{fig:a6_U}
\end{figure*}

Starting with the $6$-node random graph depicted in Fig.~\ref{fig:a6_U}(left) alongside its minimal cycle basis, we present the average energy of an $N=6$-boson system in Fig.~\ref{fig:a6_U}(right) as a function of the on-site repulsion $U$. 

\begin{figure*}[htp]
\includegraphics[scale=0.25]{graph_arbitrary_17.png}
\includegraphics[scale=0.36]{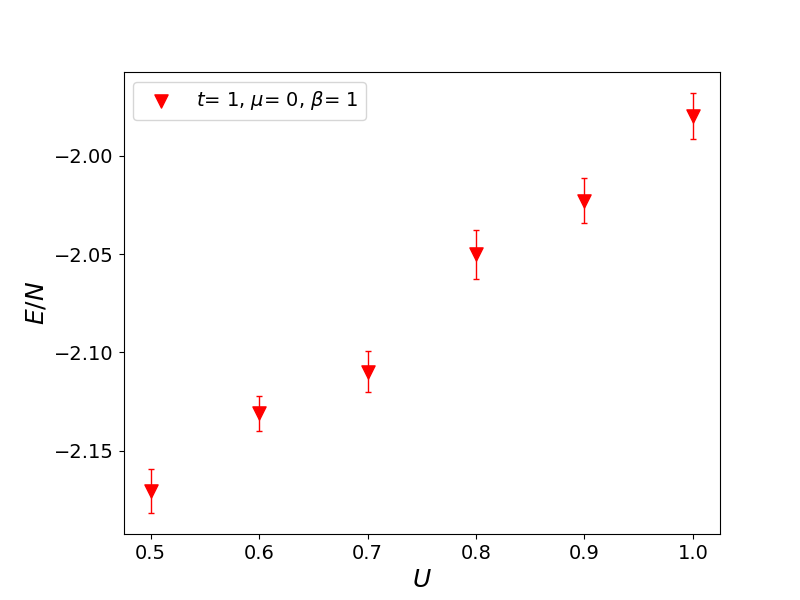}
\caption{Left: The minimal cycle basis (in red) for a $17$-node random graph.
Right: Average energy $E=\langle H \rangle$ as a function of $U$ for the graph depicted in the left panel. Here, the number of particles is $N=17$. The remaining parameters are fixed and have the following values: $t=1, \mu=0$ and $\beta=1$. }
\label{fig:a17_U}
\end{figure*}

In Fig.~\ref{fig:a17_U}(right), we show results of simulations conducted on the $17$-site graph shown in Fig.~\ref{fig:a17_U}(left). Here, we measure the total energy of the system as a function of $U$ for an $N=17$-boson system. 

\subsection{Convergence properties of the algorithm}

\begin{figure*}[htp]
	\includegraphics[scale=0.50]{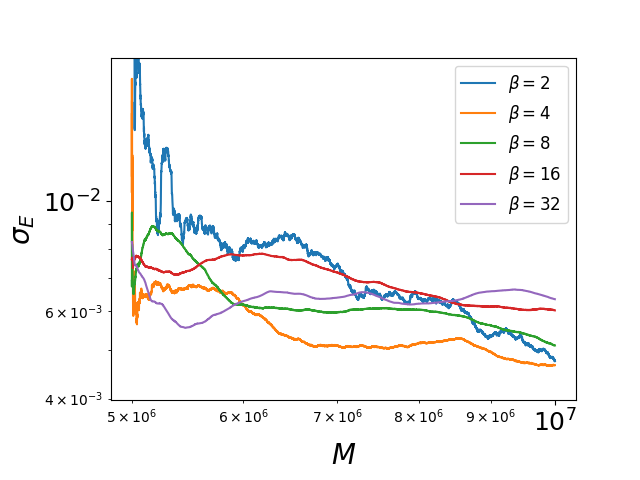}
	\includegraphics[scale=0.47]{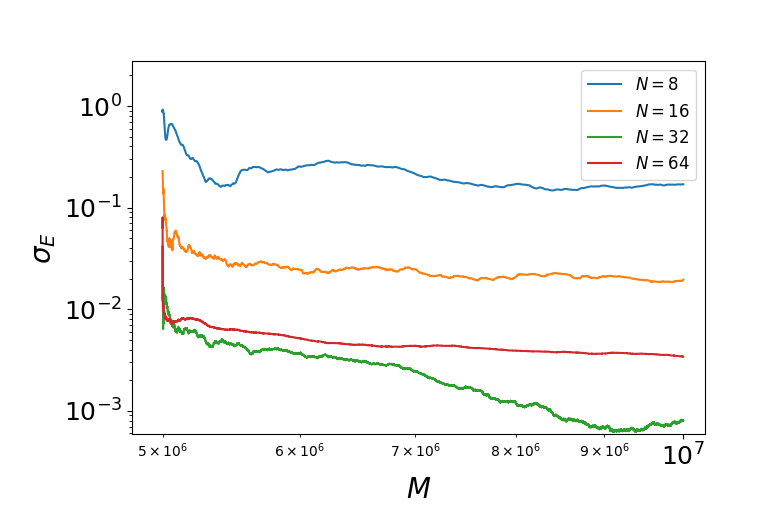}
	\caption{Left: Scaling of the error with number of QMC sweeps $M$ for different inverse-temperatures (log-log scale). Error is calculated based on the standard deviation of $10$ independent runs. We observe a power-law dependence with an average power of roughly $-2$. Right: Scaling of the error with runtime for different choices for number of particles $N$ (log-log scale).  Results shown for a $64$-boson system on an $8$ by $8$ lattice with open boundary conditions ($t=1, \mu=0, U=0.001$.)}
	\label{fig:convergence_vs_M}
\end{figure*}

We further tested the convergence properties of our algorithm by monitoring the error of the mean on an $8$ by $8$ lattice with open boundary conditions containing $64$ bosons ($t=1, \mu=0, U=0.001$) across different inverse temperatures. We plot the error against number of sweepsin Fig.~\ref{fig:convergence_vs_M} on a log-log scale. As expected, the figure indicates a power-law dependence of the error with simulation time. For each temperature, the error was calculated based on the standard deviation across $10$ independent simulation runs. 

\section{Summary and conclusions\label{sec:conc}}

We presented a quantum Monte Carlo algorithm designed to reliably simulate the Bose-Hubbard model on arbitrary graphs. We showed that a provably ergodic QMC algorithm can be devised by adapting the Permutation Matrix Representation QMC~\cite{Gupta20} augmenting it with update moves based on the minimal cycle basis of the BH graph, which can  be produced in an automated way.

To demonstrate the versatility and generality of our approach, we presented simulation results for the Bose-Hubbard model defined on regular lattices with open and periodic boundary conditions as well as on a number of irregular graphs. 

We believe that the algorithm presented in this study may become a very useful tool in the study of the equilibrium properties of Bose-Hubbard models in different dimensions and setups, which have  so far not been amenable to simulations. 

Moreover, the methods presented in this paper are readily generalizable to other types of systems, e.g., fermionic or spin systems. We aim to explore such extended techniques in future work. 

\begin{acknowledgments}
This project was supported in part by NSF award \#2210374. In addition, this material is based upon work supported
by the Defense Advanced Research Projects Agency (DARPA) under Contract No. HR001122C0063. All material, except scientific articles or papers published in scientific journals, must, in addition to any notices
or disclaimers by the Contractor, also contain the following disclaimer: Any opinions, findings and conclusions or recommendations expressed in this material are those of the author(s) and do not necessarily reflect the views of the Defense Advanced Research Projects Agency (DARPA).
\end{acknowledgments}

\bibliography{bose_hubbard_qmc}

\begin{thebibliography}{57}%
\makeatletter
\providecommand \@ifxundefined [1]{%
 \@ifx{#1\undefined}
}%
\providecommand \@ifnum [1]{%
 \ifnum #1\expandafter \@firstoftwo
 \else \expandafter \@secondoftwo
 \fi
}%
\providecommand \@ifx [1]{%
 \ifx #1\expandafter \@firstoftwo
 \else \expandafter \@secondoftwo
 \fi
}%
\providecommand \natexlab [1]{#1}%
\providecommand \enquote  [1]{``#1''}%
\providecommand \bibnamefont  [1]{#1}%
\providecommand \bibfnamefont [1]{#1}%
\providecommand \citenamefont [1]{#1}%
\providecommand \href@noop [0]{\@secondoftwo}%
\providecommand \href [0]{\begingroup \@sanitize@url \@href}%
\providecommand \@href[1]{\@@startlink{#1}\@@href}%
\providecommand \@@href[1]{\endgroup#1\@@endlink}%
\providecommand \@sanitize@url [0]{\catcode `\\12\catcode `\$12\catcode
  `\&12\catcode `\#12\catcode `\^12\catcode `\_12\catcode `\%12\relax}%
\providecommand \@@startlink[1]{}%
\providecommand \@@endlink[0]{}%
\providecommand \url  [0]{\begingroup\@sanitize@url \@url }%
\providecommand \@url [1]{\endgroup\@href {#1}{\urlprefix }}%
\providecommand \urlprefix  [0]{URL }%
\providecommand \Eprint [0]{\href }%
\providecommand \doibase [0]{https://doi.org/}%
\providecommand \selectlanguage [0]{\@gobble}%
\providecommand \bibinfo  [0]{\@secondoftwo}%
\providecommand \bibfield  [0]{\@secondoftwo}%
\providecommand \translation [1]{[#1]}%
\providecommand \BibitemOpen [0]{}%
\providecommand \bibitemStop [0]{}%
\providecommand \bibitemNoStop [0]{.\EOS\space}%
\providecommand \EOS [0]{\spacefactor3000\relax}%
\providecommand \BibitemShut  [1]{\csname bibitem#1\endcsname}%
\let\auto@bib@innerbib\@empty
\bibitem [{\citenamefont {Lewenstein}\ \emph {et~al.}(2012)\citenamefont
  {Lewenstein}, \citenamefont {Sanpera},\ and\ \citenamefont
  {Ahufinger}}]{Lewenstein12}%
  \BibitemOpen
  \bibfield  {author} {\bibinfo {author} {\bibfnamefont {M.}~\bibnamefont
  {Lewenstein}}, \bibinfo {author} {\bibfnamefont {A.}~\bibnamefont
  {Sanpera}},\ and\ \bibinfo {author} {\bibfnamefont {V.}~\bibnamefont
  {Ahufinger}},\ }\href@noop {} {\emph {\bibinfo {title} {Ultracold Atoms in
  Optical Lattices: Simulating quantum many-body systems}}}\ (\bibinfo
  {publisher} {OUP Oxford},\ \bibinfo {year} {2012})\BibitemShut {NoStop}%
\bibitem [{\citenamefont {Fisher}\ \emph {et~al.}(1989)\citenamefont {Fisher},
  \citenamefont {Weichman}, \citenamefont {Grinstein},\ and\ \citenamefont
  {Fisher}}]{Fisher89}%
  \BibitemOpen
  \bibfield  {author} {\bibinfo {author} {\bibfnamefont {M.~P.}\ \bibnamefont
  {Fisher}}, \bibinfo {author} {\bibfnamefont {P.~B.}\ \bibnamefont
  {Weichman}}, \bibinfo {author} {\bibfnamefont {G.}~\bibnamefont
  {Grinstein}},\ and\ \bibinfo {author} {\bibfnamefont {D.~S.}\ \bibnamefont
  {Fisher}},\ }\bibfield  {title} {\bibinfo {title} {Boson localization and the
  superfluid-insulator transition},\ }\href@noop {} {\bibfield  {journal}
  {\bibinfo  {journal} {Physical Review B}\ }\textbf {\bibinfo {volume} {40}},\
  \bibinfo {pages} {546} (\bibinfo {year} {1989})}\BibitemShut {NoStop}%
\bibitem [{\citenamefont {Jaksch}\ and\ \citenamefont
  {Zoller}(2005)}]{Jaksch05}%
  \BibitemOpen
  \bibfield  {author} {\bibinfo {author} {\bibfnamefont {D.}~\bibnamefont
  {Jaksch}}\ and\ \bibinfo {author} {\bibfnamefont {P.}~\bibnamefont
  {Zoller}},\ }\bibfield  {title} {\bibinfo {title} {The cold atom hubbard
  toolbox},\ }\href@noop {} {\bibfield  {journal} {\bibinfo  {journal} {Annals
  of physics}\ }\textbf {\bibinfo {volume} {315}},\ \bibinfo {pages} {52}
  (\bibinfo {year} {2005})}\BibitemShut {NoStop}%
\bibitem [{\citenamefont {Giamarchi}\ \emph {et~al.}(2008)\citenamefont
  {Giamarchi}, \citenamefont {R{\"u}egg},\ and\ \citenamefont
  {Tchernyshyov}}]{Giamarchi08}%
  \BibitemOpen
  \bibfield  {author} {\bibinfo {author} {\bibfnamefont {T.}~\bibnamefont
  {Giamarchi}}, \bibinfo {author} {\bibfnamefont {C.}~\bibnamefont
  {R{\"u}egg}},\ and\ \bibinfo {author} {\bibfnamefont {O.}~\bibnamefont
  {Tchernyshyov}},\ }\bibfield  {title} {\bibinfo {title} {Bose--einstein
  condensation in magnetic insulators},\ }\href@noop {} {\bibfield  {journal}
  {\bibinfo  {journal} {Nature Physics}\ }\textbf {\bibinfo {volume} {4}},\
  \bibinfo {pages} {198} (\bibinfo {year} {2008})}\BibitemShut {NoStop}%
\bibitem [{\citenamefont {Batrouni}\ \emph {et~al.}(1995)\citenamefont
  {Batrouni}, \citenamefont {Scalettar}, \citenamefont {Zimanyi},\ and\
  \citenamefont {Kampf}}]{Batrouni95}%
  \BibitemOpen
  \bibfield  {author} {\bibinfo {author} {\bibfnamefont {G.~G.}\ \bibnamefont
  {Batrouni}}, \bibinfo {author} {\bibfnamefont {R.~T.}\ \bibnamefont
  {Scalettar}}, \bibinfo {author} {\bibfnamefont {G.~T.}\ \bibnamefont
  {Zimanyi}},\ and\ \bibinfo {author} {\bibfnamefont {A.~P.}\ \bibnamefont
  {Kampf}},\ }\bibfield  {title} {\bibinfo {title} {Supersolids in the
  bose-hubbard hamiltonian},\ }\href
  {https://doi.org/10.1103/PhysRevLett.74.2527} {\bibfield  {journal} {\bibinfo
   {journal} {Phys. Rev. Lett.}\ }\textbf {\bibinfo {volume} {74}},\ \bibinfo
  {pages} {2527} (\bibinfo {year} {1995})}\BibitemShut {NoStop}%
\bibitem [{\citenamefont {H\'ebert}\ \emph {et~al.}(2001)\citenamefont
  {H\'ebert}, \citenamefont {Batrouni}, \citenamefont {Scalettar},
  \citenamefont {Schmid}, \citenamefont {Troyer},\ and\ \citenamefont
  {Dorneich}}]{Hebert01}%
  \BibitemOpen
  \bibfield  {author} {\bibinfo {author} {\bibfnamefont {F.}~\bibnamefont
  {H\'ebert}}, \bibinfo {author} {\bibfnamefont {G.~G.}\ \bibnamefont
  {Batrouni}}, \bibinfo {author} {\bibfnamefont {R.~T.}\ \bibnamefont
  {Scalettar}}, \bibinfo {author} {\bibfnamefont {G.}~\bibnamefont {Schmid}},
  \bibinfo {author} {\bibfnamefont {M.}~\bibnamefont {Troyer}},\ and\ \bibinfo
  {author} {\bibfnamefont {A.}~\bibnamefont {Dorneich}},\ }\bibfield  {title}
  {\bibinfo {title} {Quantum phase transitions in the two-dimensional hardcore
  boson model},\ }\href {https://doi.org/10.1103/PhysRevB.65.014513} {\bibfield
   {journal} {\bibinfo  {journal} {Phys. Rev. B}\ }\textbf {\bibinfo {volume}
  {65}},\ \bibinfo {pages} {014513} (\bibinfo {year} {2001})}\BibitemShut
  {NoStop}%
\bibitem [{\citenamefont {Sengupta}\ \emph {et~al.}(2005)\citenamefont
  {Sengupta}, \citenamefont {Pryadko}, \citenamefont {Alet}, \citenamefont
  {Troyer},\ and\ \citenamefont {Schmid}}]{Sengupta05}%
  \BibitemOpen
  \bibfield  {author} {\bibinfo {author} {\bibfnamefont {P.}~\bibnamefont
  {Sengupta}}, \bibinfo {author} {\bibfnamefont {L.~P.}\ \bibnamefont
  {Pryadko}}, \bibinfo {author} {\bibfnamefont {F.}~\bibnamefont {Alet}},
  \bibinfo {author} {\bibfnamefont {M.}~\bibnamefont {Troyer}},\ and\ \bibinfo
  {author} {\bibfnamefont {G.}~\bibnamefont {Schmid}},\ }\bibfield  {title}
  {\bibinfo {title} {Supersolids versus phase separation in two-dimensional
  lattice bosons},\ }\href {https://doi.org/10.1103/PhysRevLett.94.207202}
  {\bibfield  {journal} {\bibinfo  {journal} {Phys. Rev. Lett.}\ }\textbf
  {\bibinfo {volume} {94}},\ \bibinfo {pages} {207202} (\bibinfo {year}
  {2005})}\BibitemShut {NoStop}%
\bibitem [{\citenamefont {Heidarian}\ and\ \citenamefont
  {Damle}(2005)}]{Heidarian05}%
  \BibitemOpen
  \bibfield  {author} {\bibinfo {author} {\bibfnamefont {D.}~\bibnamefont
  {Heidarian}}\ and\ \bibinfo {author} {\bibfnamefont {K.}~\bibnamefont
  {Damle}},\ }\bibfield  {title} {\bibinfo {title} {Persistent supersolid phase
  of hard-core bosons on the triangular lattice},\ }\href
  {https://doi.org/10.1103/PhysRevLett.95.127206} {\bibfield  {journal}
  {\bibinfo  {journal} {Phys. Rev. Lett.}\ }\textbf {\bibinfo {volume} {95}},\
  \bibinfo {pages} {127206} (\bibinfo {year} {2005})}\BibitemShut {NoStop}%
\bibitem [{\citenamefont {Wessel}\ and\ \citenamefont
  {Troyer}(2005)}]{Wessel05}%
  \BibitemOpen
  \bibfield  {author} {\bibinfo {author} {\bibfnamefont {S.}~\bibnamefont
  {Wessel}}\ and\ \bibinfo {author} {\bibfnamefont {M.}~\bibnamefont
  {Troyer}},\ }\bibfield  {title} {\bibinfo {title} {Supersolid hard-core
  bosons on the triangular lattice},\ }\href
  {https://doi.org/10.1103/PhysRevLett.95.127205} {\bibfield  {journal}
  {\bibinfo  {journal} {Phys. Rev. Lett.}\ }\textbf {\bibinfo {volume} {95}},\
  \bibinfo {pages} {127205} (\bibinfo {year} {2005})}\BibitemShut {NoStop}%
\bibitem [{\citenamefont {Boninsegni}\ and\ \citenamefont
  {Prokof'ev}(2005)}]{Boninsegni05}%
  \BibitemOpen
  \bibfield  {author} {\bibinfo {author} {\bibfnamefont {M.}~\bibnamefont
  {Boninsegni}}\ and\ \bibinfo {author} {\bibfnamefont {N.}~\bibnamefont
  {Prokof'ev}},\ }\bibfield  {title} {\bibinfo {title} {Supersolid phase of
  hard-core bosons on a triangular lattice},\ }\href
  {https://doi.org/10.1103/PhysRevLett.95.237204} {\bibfield  {journal}
  {\bibinfo  {journal} {Phys. Rev. Lett.}\ }\textbf {\bibinfo {volume} {95}},\
  \bibinfo {pages} {237204} (\bibinfo {year} {2005})}\BibitemShut {NoStop}%
\bibitem [{\citenamefont {Gan}\ \emph {et~al.}(2007{\natexlab{a}})\citenamefont
  {Gan}, \citenamefont {Wen},\ and\ \citenamefont {Yu}}]{Gan07}%
  \BibitemOpen
  \bibfield  {author} {\bibinfo {author} {\bibfnamefont {J.-Y.}\ \bibnamefont
  {Gan}}, \bibinfo {author} {\bibfnamefont {Y.-C.}\ \bibnamefont {Wen}},\ and\
  \bibinfo {author} {\bibfnamefont {Y.}~\bibnamefont {Yu}},\ }\bibfield
  {title} {\bibinfo {title} {Supersolidity and phase diagram of soft-core
  bosons on a triangular lattice},\ }\href
  {https://doi.org/10.1103/PhysRevB.75.094501} {\bibfield  {journal} {\bibinfo
  {journal} {Phys. Rev. B}\ }\textbf {\bibinfo {volume} {75}},\ \bibinfo
  {pages} {094501} (\bibinfo {year} {2007}{\natexlab{a}})}\BibitemShut
  {NoStop}%
\bibitem [{\citenamefont {Bogner}\ \emph {et~al.}(2019)\citenamefont {Bogner},
  \citenamefont {De~Daniloff},\ and\ \citenamefont {Rieger}}]{Bogner19}%
  \BibitemOpen
  \bibfield  {author} {\bibinfo {author} {\bibfnamefont {B.}~\bibnamefont
  {Bogner}}, \bibinfo {author} {\bibfnamefont {C.}~\bibnamefont
  {De~Daniloff}},\ and\ \bibinfo {author} {\bibfnamefont {H.}~\bibnamefont
  {Rieger}},\ }\bibfield  {title} {\bibinfo {title} {Variational {Monte-Carlo}
  study of the extended {Bose-Hubbard} model with short- and infinite-range
  interactions},\ }\href@noop {} {\bibfield  {journal} {\bibinfo  {journal}
  {The European Physical Journal B}\ }\textbf {\bibinfo {volume} {92}},\
  \bibinfo {pages} {111} (\bibinfo {year} {2019})}\BibitemShut {NoStop}%
\bibitem [{\citenamefont {Krauth}\ \emph {et~al.}(1991)\citenamefont {Krauth},
  \citenamefont {Trivedi},\ and\ \citenamefont {Ceperley}}]{Krauth91}%
  \BibitemOpen
  \bibfield  {author} {\bibinfo {author} {\bibfnamefont {W.}~\bibnamefont
  {Krauth}}, \bibinfo {author} {\bibfnamefont {N.}~\bibnamefont {Trivedi}},\
  and\ \bibinfo {author} {\bibfnamefont {D.}~\bibnamefont {Ceperley}},\
  }\bibfield  {title} {\bibinfo {title} {Superfluid-insulator transition in
  disordered boson systems},\ }\href
  {https://doi.org/10.1103/PhysRevLett.67.2307} {\bibfield  {journal} {\bibinfo
   {journal} {Phys. Rev. Lett.}\ }\textbf {\bibinfo {volume} {67}},\ \bibinfo
  {pages} {2307} (\bibinfo {year} {1991})}\BibitemShut {NoStop}%
\bibitem [{\citenamefont {Krauth}\ and\ \citenamefont
  {Trivedi}(1991)}]{Krauth91-2}%
  \BibitemOpen
  \bibfield  {author} {\bibinfo {author} {\bibfnamefont {W.}~\bibnamefont
  {Krauth}}\ and\ \bibinfo {author} {\bibfnamefont {N.}~\bibnamefont
  {Trivedi}},\ }\bibfield  {title} {\bibinfo {title} {Mott and superfluid
  transitions in a strongly interacting lattice boson system},\ }\href@noop {}
  {\bibfield  {journal} {\bibinfo  {journal} {EPL (Europhysics Letters)}\
  }\textbf {\bibinfo {volume} {14}},\ \bibinfo {pages} {627} (\bibinfo {year}
  {1991})}\BibitemShut {NoStop}%
\bibitem [{\citenamefont {Kisker}\ and\ \citenamefont
  {Rieger}(1997)}]{Kisker97}%
  \BibitemOpen
  \bibfield  {author} {\bibinfo {author} {\bibfnamefont {J.}~\bibnamefont
  {Kisker}}\ and\ \bibinfo {author} {\bibfnamefont {H.}~\bibnamefont
  {Rieger}},\ }\bibfield  {title} {\bibinfo {title} {Bose-glass and
  mott-insulator phase in the disordered boson hubbard model},\ }\href
  {https://doi.org/10.1103/PhysRevB.55.R11981} {\bibfield  {journal} {\bibinfo
  {journal} {Phys. Rev. B}\ }\textbf {\bibinfo {volume} {55}},\ \bibinfo
  {pages} {R11981} (\bibinfo {year} {1997})}\BibitemShut {NoStop}%
\bibitem [{\citenamefont {Melko}\ \emph {et~al.}(2005)\citenamefont {Melko},
  \citenamefont {Paramekanti}, \citenamefont {Burkov}, \citenamefont
  {Vishwanath}, \citenamefont {Sheng},\ and\ \citenamefont
  {Balents}}]{Melko05}%
  \BibitemOpen
  \bibfield  {author} {\bibinfo {author} {\bibfnamefont {R.~G.}\ \bibnamefont
  {Melko}}, \bibinfo {author} {\bibfnamefont {A.}~\bibnamefont {Paramekanti}},
  \bibinfo {author} {\bibfnamefont {A.~A.}\ \bibnamefont {Burkov}}, \bibinfo
  {author} {\bibfnamefont {A.}~\bibnamefont {Vishwanath}}, \bibinfo {author}
  {\bibfnamefont {D.~N.}\ \bibnamefont {Sheng}},\ and\ \bibinfo {author}
  {\bibfnamefont {L.}~\bibnamefont {Balents}},\ }\bibfield  {title} {\bibinfo
  {title} {Supersolid order from disorder: Hard-core bosons on the triangular
  lattice},\ }\href {https://doi.org/10.1103/PhysRevLett.95.127207} {\bibfield
  {journal} {\bibinfo  {journal} {Phys. Rev. Lett.}\ }\textbf {\bibinfo
  {volume} {95}},\ \bibinfo {pages} {127207} (\bibinfo {year}
  {2005})}\BibitemShut {NoStop}%
\bibitem [{\citenamefont {Yokoyama}\ \emph {et~al.}(2011)\citenamefont
  {Yokoyama}, \citenamefont {Miyagawa},\ and\ \citenamefont
  {Ogata}}]{Yokoyama11}%
  \BibitemOpen
  \bibfield  {author} {\bibinfo {author} {\bibfnamefont {H.}~\bibnamefont
  {Yokoyama}}, \bibinfo {author} {\bibfnamefont {T.}~\bibnamefont {Miyagawa}},\
  and\ \bibinfo {author} {\bibfnamefont {M.}~\bibnamefont {Ogata}},\ }\bibfield
   {title} {\bibinfo {title} {Effect of doublon–holon binding on mott
  transition–variational monte carlo study of two-dimensional bose hubbard
  models},\ }\href {https://doi.org/10.1143/JPSJ.80.084607} {\bibfield
  {journal} {\bibinfo  {journal} {Journal of the Physical Society of Japan}\
  }\textbf {\bibinfo {volume} {80}},\ \bibinfo {pages} {084607} (\bibinfo
  {year} {2011})},\ \Eprint
  {https://arxiv.org/abs/https://doi.org/10.1143/JPSJ.80.084607}
  {https://doi.org/10.1143/JPSJ.80.084607} \BibitemShut {NoStop}%
\bibitem [{\citenamefont {Scalettar}\ \emph {et~al.}(1991)\citenamefont
  {Scalettar}, \citenamefont {Batrouni},\ and\ \citenamefont
  {Zimanyi}}]{Scalettar91}%
  \BibitemOpen
  \bibfield  {author} {\bibinfo {author} {\bibfnamefont {R.~T.}\ \bibnamefont
  {Scalettar}}, \bibinfo {author} {\bibfnamefont {G.~G.}\ \bibnamefont
  {Batrouni}},\ and\ \bibinfo {author} {\bibfnamefont {G.~T.}\ \bibnamefont
  {Zimanyi}},\ }\bibfield  {title} {\bibinfo {title} {Localization in
  interacting, disordered, bose systems},\ }\href@noop {} {\bibfield  {journal}
  {\bibinfo  {journal} {Physical review letters}\ }\textbf {\bibinfo {volume}
  {66}},\ \bibinfo {pages} {3144} (\bibinfo {year} {1991})}\BibitemShut
  {NoStop}%
\bibitem [{\citenamefont {Sengupta}\ and\ \citenamefont
  {Haas}(2007)}]{Sengupta07}%
  \BibitemOpen
  \bibfield  {author} {\bibinfo {author} {\bibfnamefont {P.}~\bibnamefont
  {Sengupta}}\ and\ \bibinfo {author} {\bibfnamefont {S.}~\bibnamefont
  {Haas}},\ }\bibfield  {title} {\bibinfo {title} {Quantum glass phases in the
  disordered bose-hubbard model},\ }\href
  {https://doi.org/10.1103/PhysRevLett.99.050403} {\bibfield  {journal}
  {\bibinfo  {journal} {Phys. Rev. Lett.}\ }\textbf {\bibinfo {volume} {99}},\
  \bibinfo {pages} {050403} (\bibinfo {year} {2007})}\BibitemShut {NoStop}%
\bibitem [{\citenamefont {{Prokof'ev}}(2003)}]{Prokofev03}%
  \BibitemOpen
  \bibfield  {author} {\bibinfo {author} {\bibfnamefont {N.}~\bibnamefont
  {{Prokof'ev}}},\ }\bibfield  {title} {\bibinfo {title} {{Revealing
  Superfluid--Mott-Insulator Transition in an Optical Lattice}},\ }in\
  \href@noop {} {\emph {\bibinfo {booktitle} {APS March Meeting Abstracts}}},\
  \bibinfo {series} {APS Meeting Abstracts}, Vol.\ \bibinfo {volume} {2003}\
  (\bibinfo {year} {2003})\ p.\ \bibinfo {pages} {H4.002}\BibitemShut {NoStop}%
\bibitem [{\citenamefont {Wessel}\ \emph {et~al.}(2004)\citenamefont {Wessel},
  \citenamefont {Alet}, \citenamefont {Troyer},\ and\ \citenamefont
  {Batrouni}}]{Wessel04}%
  \BibitemOpen
  \bibfield  {author} {\bibinfo {author} {\bibfnamefont {S.}~\bibnamefont
  {Wessel}}, \bibinfo {author} {\bibfnamefont {F.}~\bibnamefont {Alet}},
  \bibinfo {author} {\bibfnamefont {M.}~\bibnamefont {Troyer}},\ and\ \bibinfo
  {author} {\bibfnamefont {G.~G.}\ \bibnamefont {Batrouni}},\ }\bibfield
  {title} {\bibinfo {title} {Quantum monte carlo simulations of confined
  bosonic atoms in optical lattices},\ }\href@noop {} {\bibfield  {journal}
  {\bibinfo  {journal} {Physical Review A}\ }\textbf {\bibinfo {volume} {70}},\
  \bibinfo {pages} {053615} (\bibinfo {year} {2004})}\BibitemShut {NoStop}%
\bibitem [{\citenamefont {Wessel}\ \emph {et~al.}(2005)\citenamefont {Wessel},
  \citenamefont {Alet}, \citenamefont {Trebst}, \citenamefont {Leumann},
  \citenamefont {Troyer},\ and\ \citenamefont {George~Batrouni}}]{Wessel05-2}%
  \BibitemOpen
  \bibfield  {author} {\bibinfo {author} {\bibfnamefont {S.}~\bibnamefont
  {Wessel}}, \bibinfo {author} {\bibfnamefont {F.}~\bibnamefont {Alet}},
  \bibinfo {author} {\bibfnamefont {S.}~\bibnamefont {Trebst}}, \bibinfo
  {author} {\bibfnamefont {D.}~\bibnamefont {Leumann}}, \bibinfo {author}
  {\bibfnamefont {M.}~\bibnamefont {Troyer}},\ and\ \bibinfo {author}
  {\bibfnamefont {G.}~\bibnamefont {George~Batrouni}},\ }\bibfield  {title}
  {\bibinfo {title} {Bosons in optical lattices--from the mott transition to
  the tonks--girardeau gas},\ }\href@noop {} {\bibfield  {journal} {\bibinfo
  {journal} {Journal of the Physical Society of Japan}\ }\textbf {\bibinfo
  {volume} {74}},\ \bibinfo {pages} {10} (\bibinfo {year} {2005})}\BibitemShut
  {NoStop}%
\bibitem [{\citenamefont {Pollet}\ \emph {et~al.}(2008)\citenamefont {Pollet},
  \citenamefont {Kollath}, \citenamefont {Houcke},\ and\ \citenamefont
  {Troyer}}]{Pollet_2008}%
  \BibitemOpen
  \bibfield  {author} {\bibinfo {author} {\bibfnamefont {L.}~\bibnamefont
  {Pollet}}, \bibinfo {author} {\bibfnamefont {C.}~\bibnamefont {Kollath}},
  \bibinfo {author} {\bibfnamefont {K.~V.}\ \bibnamefont {Houcke}},\ and\
  \bibinfo {author} {\bibfnamefont {M.}~\bibnamefont {Troyer}},\ }\bibfield
  {title} {\bibinfo {title} {Temperature changes when adiabatically ramping up
  an optical lattice},\ }\href {https://doi.org/10.1088/1367-2630/10/6/065001}
  {\bibfield  {journal} {\bibinfo  {journal} {New Journal of Physics}\ }\textbf
  {\bibinfo {volume} {10}},\ \bibinfo {pages} {065001} (\bibinfo {year}
  {2008})}\BibitemShut {NoStop}%
\bibitem [{\citenamefont {Gan}\ \emph {et~al.}(2007{\natexlab{b}})\citenamefont
  {Gan}, \citenamefont {Wen}, \citenamefont {Ye}, \citenamefont {Li},
  \citenamefont {Yang},\ and\ \citenamefont {Yu}}]{Gan07-2}%
  \BibitemOpen
  \bibfield  {author} {\bibinfo {author} {\bibfnamefont {J.~Y.}\ \bibnamefont
  {Gan}}, \bibinfo {author} {\bibfnamefont {Y.~C.}\ \bibnamefont {Wen}},
  \bibinfo {author} {\bibfnamefont {J.}~\bibnamefont {Ye}}, \bibinfo {author}
  {\bibfnamefont {T.}~\bibnamefont {Li}}, \bibinfo {author} {\bibfnamefont
  {S.-J.}\ \bibnamefont {Yang}},\ and\ \bibinfo {author} {\bibfnamefont
  {Y.}~\bibnamefont {Yu}},\ }\bibfield  {title} {\bibinfo {title} {Extended
  bose-hubbard model on a honeycomb lattice},\ }\href@noop {} {\bibfield
  {journal} {\bibinfo  {journal} {Physical Review B}\ }\textbf {\bibinfo
  {volume} {75}},\ \bibinfo {pages} {214509} (\bibinfo {year}
  {2007}{\natexlab{b}})}\BibitemShut {NoStop}%
\bibitem [{\citenamefont {Kawaki}\ \emph {et~al.}(2017)\citenamefont {Kawaki},
  \citenamefont {Kuno},\ and\ \citenamefont {Ichinose}}]{Kawaki17}%
  \BibitemOpen
  \bibfield  {author} {\bibinfo {author} {\bibfnamefont {K.}~\bibnamefont
  {Kawaki}}, \bibinfo {author} {\bibfnamefont {Y.}~\bibnamefont {Kuno}},\ and\
  \bibinfo {author} {\bibfnamefont {I.}~\bibnamefont {Ichinose}},\ }\bibfield
  {title} {\bibinfo {title} {Phase diagrams of the extended bose-hubbard model
  in one dimension by monte-carlo simulation with the help of a
  stochastic-series expansion},\ }\href
  {https://doi.org/10.1103/PhysRevB.95.195101} {\bibfield  {journal} {\bibinfo
  {journal} {Phys. Rev. B}\ }\textbf {\bibinfo {volume} {95}},\ \bibinfo
  {pages} {195101} (\bibinfo {year} {2017})}\BibitemShut {NoStop}%
\bibitem [{\citenamefont {Sandvik}(1992)}]{sandvik:92}%
  \BibitemOpen
  \bibfield  {author} {\bibinfo {author} {\bibfnamefont {A.~W.}\ \bibnamefont
  {Sandvik}},\ }\bibfield  {title} {\bibinfo {title} {A generalization of
  {H}andscomb's quantum {M}onte {C}arlo scheme --- {A}pplication to the 1-{D}
  {H}ubbard model},\ }\href@noop {} {\bibfield  {journal} {\bibinfo  {journal}
  {J. Phys, A}\ }\textbf {\bibinfo {volume} {25}},\ \bibinfo {pages} {3667}
  (\bibinfo {year} {1992})}\BibitemShut {NoStop}%
\bibitem [{\citenamefont {Sandvik}(1999)}]{sandvik:99}%
  \BibitemOpen
  \bibfield  {author} {\bibinfo {author} {\bibfnamefont {A.~W.}\ \bibnamefont
  {Sandvik}},\ }\bibfield  {title} {\bibinfo {title} {Stochastic series
  expansion method with operator-loop update},\ }\href
  {https://doi.org/10.1103/PhysRevB.59.R14157} {\bibfield  {journal} {\bibinfo
  {journal} {Phys. Rev. B}\ }\textbf {\bibinfo {volume} {59}},\ \bibinfo
  {pages} {R14157} (\bibinfo {year} {1999})}\BibitemShut {NoStop}%
\bibitem [{\citenamefont {Sandvik}(2005)}]{sandvik:05}%
  \BibitemOpen
  \bibfield  {author} {\bibinfo {author} {\bibfnamefont {A.~W.}\ \bibnamefont
  {Sandvik}},\ }\bibfield  {title} {\bibinfo {title} {Ground state projection
  of quantum spin systems in the valence-bond basis},\ }\href
  {https://doi.org/10.1103/PhysRevLett.95.207203} {\bibfield  {journal}
  {\bibinfo  {journal} {Phys. Rev. Lett.}\ }\textbf {\bibinfo {volume} {95}},\
  \bibinfo {pages} {207203} (\bibinfo {year} {2005})}\BibitemShut {NoStop}%
\bibitem [{\citenamefont {Sandvik}\ and\ \citenamefont
  {Evertz}(2010)}]{sandvik:10}%
  \BibitemOpen
  \bibfield  {author} {\bibinfo {author} {\bibfnamefont {A.~W.}\ \bibnamefont
  {Sandvik}}\ and\ \bibinfo {author} {\bibfnamefont {H.~G.}\ \bibnamefont
  {Evertz}},\ }\bibfield  {title} {\bibinfo {title} {Loop updates for
  variational and projector quantum monte carlo simulations in the valence-bond
  basis},\ }\href {https://doi.org/10.1103/PhysRevB.82.024407} {\bibfield
  {journal} {\bibinfo  {journal} {Phys. Rev. B}\ }\textbf {\bibinfo {volume}
  {82}},\ \bibinfo {pages} {024407} (\bibinfo {year} {2010})}\BibitemShut
  {NoStop}%
\bibitem [{\citenamefont {Isakov}\ \emph {et~al.}(2006)\citenamefont {Isakov},
  \citenamefont {Wessel}, \citenamefont {Melko}, \citenamefont {Sengupta},\
  and\ \citenamefont {Kim}}]{Isakov06}%
  \BibitemOpen
  \bibfield  {author} {\bibinfo {author} {\bibfnamefont {S.~V.}\ \bibnamefont
  {Isakov}}, \bibinfo {author} {\bibfnamefont {S.}~\bibnamefont {Wessel}},
  \bibinfo {author} {\bibfnamefont {R.~G.}\ \bibnamefont {Melko}}, \bibinfo
  {author} {\bibfnamefont {K.}~\bibnamefont {Sengupta}},\ and\ \bibinfo
  {author} {\bibfnamefont {Y.~B.}\ \bibnamefont {Kim}},\ }\bibfield  {title}
  {\bibinfo {title} {Hard-core bosons on the kagome lattice: Valence-bond
  solids and their quantum melting},\ }\href
  {https://doi.org/10.1103/PhysRevLett.97.147202} {\bibfield  {journal}
  {\bibinfo  {journal} {Phys. Rev. Lett.}\ }\textbf {\bibinfo {volume} {97}},\
  \bibinfo {pages} {147202} (\bibinfo {year} {2006})}\BibitemShut {NoStop}%
\bibitem [{\citenamefont {Isakov}\ \emph {et~al.}(2009)\citenamefont {Isakov},
  \citenamefont {Sengupta},\ and\ \citenamefont {Kim}}]{Isakov09}%
  \BibitemOpen
  \bibfield  {author} {\bibinfo {author} {\bibfnamefont {S.~V.}\ \bibnamefont
  {Isakov}}, \bibinfo {author} {\bibfnamefont {K.}~\bibnamefont {Sengupta}},\
  and\ \bibinfo {author} {\bibfnamefont {Y.~B.}\ \bibnamefont {Kim}},\
  }\bibfield  {title} {\bibinfo {title} {Bose-hubbard model on a star
  lattice},\ }\href {https://doi.org/10.1103/PhysRevB.80.214503} {\bibfield
  {journal} {\bibinfo  {journal} {Phys. Rev. B}\ }\textbf {\bibinfo {volume}
  {80}},\ \bibinfo {pages} {214503} (\bibinfo {year} {2009})}\BibitemShut
  {NoStop}%
\bibitem [{\citenamefont {Pippan}\ \emph {et~al.}(2009)\citenamefont {Pippan},
  \citenamefont {Evertz},\ and\ \citenamefont
  {Hohenadler}}]{PhysRevA.80.033612}%
  \BibitemOpen
  \bibfield  {author} {\bibinfo {author} {\bibfnamefont {P.}~\bibnamefont
  {Pippan}}, \bibinfo {author} {\bibfnamefont {H.~G.}\ \bibnamefont {Evertz}},\
  and\ \bibinfo {author} {\bibfnamefont {M.}~\bibnamefont {Hohenadler}},\
  }\bibfield  {title} {\bibinfo {title} {Excitation spectra of strongly
  correlated lattice bosons and polaritons},\ }\href
  {https://doi.org/10.1103/PhysRevA.80.033612} {\bibfield  {journal} {\bibinfo
  {journal} {Phys. Rev. A}\ }\textbf {\bibinfo {volume} {80}},\ \bibinfo
  {pages} {033612} (\bibinfo {year} {2009})}\BibitemShut {NoStop}%
\bibitem [{\citenamefont {Batrouni}\ and\ \citenamefont
  {Scalettar}(1992)}]{PhysRevB.46.9051}%
  \BibitemOpen
  \bibfield  {author} {\bibinfo {author} {\bibfnamefont {G.~G.}\ \bibnamefont
  {Batrouni}}\ and\ \bibinfo {author} {\bibfnamefont {R.~T.}\ \bibnamefont
  {Scalettar}},\ }\bibfield  {title} {\bibinfo {title} {World-line quantum
  monte carlo algorithm for a one-dimensional bose model},\ }\href
  {https://doi.org/10.1103/PhysRevB.46.9051} {\bibfield  {journal} {\bibinfo
  {journal} {Phys. Rev. B}\ }\textbf {\bibinfo {volume} {46}},\ \bibinfo
  {pages} {9051} (\bibinfo {year} {1992})}\BibitemShut {NoStop}%
\bibitem [{\citenamefont {van Otterlo}\ and\ \citenamefont
  {Wagenblast}(1994)}]{vanOtterlo94}%
  \BibitemOpen
  \bibfield  {author} {\bibinfo {author} {\bibfnamefont {A.}~\bibnamefont {van
  Otterlo}}\ and\ \bibinfo {author} {\bibfnamefont {K.-H.}\ \bibnamefont
  {Wagenblast}},\ }\bibfield  {title} {\bibinfo {title} {Coexistence of
  diagonal and off-diagonal long-range order: A monte carlo study},\ }\href
  {https://doi.org/10.1103/PhysRevLett.72.3598} {\bibfield  {journal} {\bibinfo
   {journal} {Phys. Rev. Lett.}\ }\textbf {\bibinfo {volume} {72}},\ \bibinfo
  {pages} {3598} (\bibinfo {year} {1994})}\BibitemShut {NoStop}%
\bibitem [{\citenamefont {Batrouni}\ and\ \citenamefont
  {Scalettar}(2000)}]{Batrouni00}%
  \BibitemOpen
  \bibfield  {author} {\bibinfo {author} {\bibfnamefont {G.~G.}\ \bibnamefont
  {Batrouni}}\ and\ \bibinfo {author} {\bibfnamefont {R.~T.}\ \bibnamefont
  {Scalettar}},\ }\bibfield  {title} {\bibinfo {title} {Phase separation in
  supersolids},\ }\href {https://doi.org/10.1103/PhysRevLett.84.1599}
  {\bibfield  {journal} {\bibinfo  {journal} {Phys. Rev. Lett.}\ }\textbf
  {\bibinfo {volume} {84}},\ \bibinfo {pages} {1599} (\bibinfo {year}
  {2000})}\BibitemShut {NoStop}%
\bibitem [{\citenamefont {Capogrosso-Sansone}\ \emph
  {et~al.}(2008)\citenamefont {Capogrosso-Sansone}, \citenamefont {S\"oyler},
  \citenamefont {Prokof'ev},\ and\ \citenamefont
  {Svistunov}}]{Capogrosso-Sansone08}%
  \BibitemOpen
  \bibfield  {author} {\bibinfo {author} {\bibfnamefont {B.}~\bibnamefont
  {Capogrosso-Sansone}}, \bibinfo {author} {\bibfnamefont {S.~G.}\ \bibnamefont
  {S\"oyler}}, \bibinfo {author} {\bibfnamefont {N.}~\bibnamefont
  {Prokof'ev}},\ and\ \bibinfo {author} {\bibfnamefont {B.}~\bibnamefont
  {Svistunov}},\ }\bibfield  {title} {\bibinfo {title} {Monte carlo study of
  the two-dimensional bose-hubbard model},\ }\href
  {https://doi.org/10.1103/PhysRevA.77.015602} {\bibfield  {journal} {\bibinfo
  {journal} {Phys. Rev. A}\ }\textbf {\bibinfo {volume} {77}},\ \bibinfo
  {pages} {015602} (\bibinfo {year} {2008})}\BibitemShut {NoStop}%
\bibitem [{\citenamefont {S\"oyler}\ \emph {et~al.}(2011)\citenamefont
  {S\"oyler}, \citenamefont {Kiselev}, \citenamefont {Prokof'ev},\ and\
  \citenamefont {Svistunov}}]{Soyler11}%
  \BibitemOpen
  \bibfield  {author} {\bibinfo {author} {\bibfnamefont {S.~G.}\ \bibnamefont
  {S\"oyler}}, \bibinfo {author} {\bibfnamefont {M.}~\bibnamefont {Kiselev}},
  \bibinfo {author} {\bibfnamefont {N.~V.}\ \bibnamefont {Prokof'ev}},\ and\
  \bibinfo {author} {\bibfnamefont {B.~V.}\ \bibnamefont {Svistunov}},\
  }\bibfield  {title} {\bibinfo {title} {Phase diagram of the commensurate
  two-dimensional disordered bose-hubbard model},\ }\href
  {https://doi.org/10.1103/PhysRevLett.107.185301} {\bibfield  {journal}
  {\bibinfo  {journal} {Phys. Rev. Lett.}\ }\textbf {\bibinfo {volume} {107}},\
  \bibinfo {pages} {185301} (\bibinfo {year} {2011})}\BibitemShut {NoStop}%
\bibitem [{\citenamefont {Ohgoe}\ \emph {et~al.}(2012)\citenamefont {Ohgoe},
  \citenamefont {Suzuki},\ and\ \citenamefont {Kawashima}}]{Ohgoe12}%
  \BibitemOpen
  \bibfield  {author} {\bibinfo {author} {\bibfnamefont {T.}~\bibnamefont
  {Ohgoe}}, \bibinfo {author} {\bibfnamefont {T.}~\bibnamefont {Suzuki}},\ and\
  \bibinfo {author} {\bibfnamefont {N.}~\bibnamefont {Kawashima}},\ }\bibfield
  {title} {\bibinfo {title} {Ground-state phase diagram of the two-dimensional
  extended bose-hubbard model},\ }\href
  {https://doi.org/10.1103/PhysRevB.86.054520} {\bibfield  {journal} {\bibinfo
  {journal} {Phys. Rev. B}\ }\textbf {\bibinfo {volume} {86}},\ \bibinfo
  {pages} {054520} (\bibinfo {year} {2012})}\BibitemShut {NoStop}%
\bibitem [{\citenamefont {Capogrosso-Sansone}\ \emph
  {et~al.}(2007)\citenamefont {Capogrosso-Sansone}, \citenamefont {Prokof'ev},\
  and\ \citenamefont {Svistunov}}]{Capogrosso-Sansone07}%
  \BibitemOpen
  \bibfield  {author} {\bibinfo {author} {\bibfnamefont {B.}~\bibnamefont
  {Capogrosso-Sansone}}, \bibinfo {author} {\bibfnamefont {N.~V.}\ \bibnamefont
  {Prokof'ev}},\ and\ \bibinfo {author} {\bibfnamefont {B.~V.}\ \bibnamefont
  {Svistunov}},\ }\bibfield  {title} {\bibinfo {title} {Phase diagram and
  thermodynamics of the three-dimensional bose-hubbard model},\ }\href
  {https://doi.org/10.1103/PhysRevB.75.134302} {\bibfield  {journal} {\bibinfo
  {journal} {Phys. Rev. B}\ }\textbf {\bibinfo {volume} {75}},\ \bibinfo
  {pages} {134302} (\bibinfo {year} {2007})}\BibitemShut {NoStop}%
\bibitem [{\citenamefont {Anders}\ \emph {et~al.}(2010)\citenamefont {Anders},
  \citenamefont {Gull}, \citenamefont {Pollet}, \citenamefont {Troyer},\ and\
  \citenamefont {Werner}}]{Anders10}%
  \BibitemOpen
  \bibfield  {author} {\bibinfo {author} {\bibfnamefont {P.}~\bibnamefont
  {Anders}}, \bibinfo {author} {\bibfnamefont {E.}~\bibnamefont {Gull}},
  \bibinfo {author} {\bibfnamefont {L.}~\bibnamefont {Pollet}}, \bibinfo
  {author} {\bibfnamefont {M.}~\bibnamefont {Troyer}},\ and\ \bibinfo {author}
  {\bibfnamefont {P.}~\bibnamefont {Werner}},\ }\bibfield  {title} {\bibinfo
  {title} {Dynamical mean field solution of the bose-hubbard model},\ }\href
  {https://doi.org/10.1103/PhysRevLett.105.096402} {\bibfield  {journal}
  {\bibinfo  {journal} {Phys. Rev. Lett.}\ }\textbf {\bibinfo {volume} {105}},\
  \bibinfo {pages} {096402} (\bibinfo {year} {2010})}\BibitemShut {NoStop}%
\bibitem [{\citenamefont {Kato}\ and\ \citenamefont
  {Kawashima}(2009)}]{Kato09}%
  \BibitemOpen
  \bibfield  {author} {\bibinfo {author} {\bibfnamefont {Y.}~\bibnamefont
  {Kato}}\ and\ \bibinfo {author} {\bibfnamefont {N.}~\bibnamefont
  {Kawashima}},\ }\bibfield  {title} {\bibinfo {title} {Quantum monte carlo
  method for the bose-hubbard model with harmonic confining potential},\
  }\href@noop {} {\bibfield  {journal} {\bibinfo  {journal} {Physical Review
  E}\ }\textbf {\bibinfo {volume} {79}},\ \bibinfo {pages} {021104} (\bibinfo
  {year} {2009})}\BibitemShut {NoStop}%
\bibitem [{\citenamefont {Hettiarachchilage}\ \emph {et~al.}(2013)\citenamefont
  {Hettiarachchilage}, \citenamefont {Rousseau}, \citenamefont {Tam},
  \citenamefont {Jarrell},\ and\ \citenamefont {Moreno}}]{Hettiarachchilage13}%
  \BibitemOpen
  \bibfield  {author} {\bibinfo {author} {\bibfnamefont {K.}~\bibnamefont
  {Hettiarachchilage}}, \bibinfo {author} {\bibfnamefont {V.~G.}\ \bibnamefont
  {Rousseau}}, \bibinfo {author} {\bibfnamefont {K.-M.}\ \bibnamefont {Tam}},
  \bibinfo {author} {\bibfnamefont {M.}~\bibnamefont {Jarrell}},\ and\ \bibinfo
  {author} {\bibfnamefont {J.}~\bibnamefont {Moreno}},\ }\bibfield  {title}
  {\bibinfo {title} {Phase diagram of the bose-hubbard model on a ring-shaped
  lattice with tunable weak links},\ }\href
  {https://doi.org/10.1103/PhysRevA.87.051607} {\bibfield  {journal} {\bibinfo
  {journal} {Phys. Rev. A}\ }\textbf {\bibinfo {volume} {87}},\ \bibinfo
  {pages} {051607} (\bibinfo {year} {2013})}\BibitemShut {NoStop}%
\bibitem [{\citenamefont {Gupta}\ \emph {et~al.}(2020)\citenamefont {Gupta},
  \citenamefont {Albash},\ and\ \citenamefont {Hen}}]{Gupta20}%
  \BibitemOpen
  \bibfield  {author} {\bibinfo {author} {\bibfnamefont {L.}~\bibnamefont
  {Gupta}}, \bibinfo {author} {\bibfnamefont {T.}~\bibnamefont {Albash}},\ and\
  \bibinfo {author} {\bibfnamefont {I.}~\bibnamefont {Hen}},\ }\bibfield
  {title} {\bibinfo {title} {Permutation matrix representation quantum monte
  carlo},\ }\href {https://doi.org/10.1088/1742-5468/ab9e64} {\bibfield
  {journal} {\bibinfo  {journal} {Journal of Statistical Mechanics: Theory and
  Experiment}\ }\textbf {\bibinfo {volume} {2020}},\ \bibinfo {pages} {073105}
  (\bibinfo {year} {2020})}\BibitemShut {NoStop}%
\bibitem [{\citenamefont {Berger}\ \emph {et~al.}(2004)\citenamefont {Berger},
  \citenamefont {Gritzmann},\ and\ \citenamefont {de~Vries}}]{Berger04}%
  \BibitemOpen
  \bibfield  {author} {\bibinfo {author} {\bibfnamefont {F.}~\bibnamefont
  {Berger}}, \bibinfo {author} {\bibfnamefont {P.}~\bibnamefont {Gritzmann}},\
  and\ \bibinfo {author} {\bibfnamefont {S.}~\bibnamefont {de~Vries}},\
  }\bibfield  {title} {\bibinfo {title} {Minimum cycle bases for network
  graphs},\ }\href@noop {} {\bibfield  {journal} {\bibinfo  {journal}
  {Algorithmica}\ }\textbf {\bibinfo {volume} {40}},\ \bibinfo {pages} {51}
  (\bibinfo {year} {2004})}\BibitemShut {NoStop}%
\bibitem [{\citenamefont {Uno}\ and\ \citenamefont {Satoh}(2014)}]{Uno14}%
  \BibitemOpen
  \bibfield  {author} {\bibinfo {author} {\bibfnamefont {T.}~\bibnamefont
  {Uno}}\ and\ \bibinfo {author} {\bibfnamefont {H.}~\bibnamefont {Satoh}},\
  }\bibfield  {title} {\bibinfo {title} {An efficient algorithm for enumerating
  chordless cycles and chordless paths},\ }in\ \href@noop {} {\emph {\bibinfo
  {booktitle} {Discovery Science}}},\ \bibinfo {editor} {edited by\ \bibinfo
  {editor} {\bibfnamefont {S.}~\bibnamefont {D{\v{z}}eroski}}, \bibinfo
  {editor} {\bibfnamefont {P.}~\bibnamefont {Panov}}, \bibinfo {editor}
  {\bibfnamefont {D.}~\bibnamefont {Kocev}},\ and\ \bibinfo {editor}
  {\bibfnamefont {L.}~\bibnamefont {Todorovski}}}\ (\bibinfo  {publisher}
  {Springer International Publishing},\ \bibinfo {address} {Cham},\ \bibinfo
  {year} {2014})\ pp.\ \bibinfo {pages} {313--324}\BibitemShut {NoStop}%
\bibitem [{\citenamefont {Joyner}(2008)}]{Joyner08}%
  \BibitemOpen
  \bibfield  {author} {\bibinfo {author} {\bibfnamefont {D.}~\bibnamefont
  {Joyner}},\ }\href {https://books.google.com/books?id=iM0fco-\_Ri8C} {\emph
  {\bibinfo {title} {Adventures in Group Theory: Rubik's Cube, Merlin's
  Machine, and Other Mathematical Toys}}},\ Adventures in Group Theory\
  (\bibinfo  {publisher} {Johns Hopkins University Press},\ \bibinfo {year}
  {2008})\BibitemShut {NoStop}%
\bibitem [{\citenamefont {Whittaker}\ and\ \citenamefont
  {Robinson}(1967)}]{Whittaker67}%
  \BibitemOpen
  \bibfield  {author} {\bibinfo {author} {\bibfnamefont {E.~T.}\ \bibnamefont
  {Whittaker}}\ and\ \bibinfo {author} {\bibfnamefont {G.}~\bibnamefont
  {Robinson}},\ }\href@noop {} {\emph {\bibinfo {title} {The calculus of
  observations: An introduction to numerical analysis}}}\ (\bibinfo
  {publisher} {Dover Publications},\ \bibinfo {year} {1967})\BibitemShut
  {NoStop}%
\bibitem [{\citenamefont {de~Boor}(2005)}]{de05}%
  \BibitemOpen
  \bibfield  {author} {\bibinfo {author} {\bibfnamefont {C.}~\bibnamefont
  {de~Boor}},\ }\bibfield  {title} {\bibinfo {title} {Divided differences.},\
  }\href@noop {} {\bibfield  {journal} {\bibinfo  {journal} {Surveys in
  Approximation Theory (SAT)[electronic only]}\ }\textbf {\bibinfo {volume}
  {1}},\ \bibinfo {pages} {46} (\bibinfo {year} {2005})}\BibitemShut {NoStop}%
\bibitem [{\citenamefont {Pollet}(2012)}]{Pollet_2012}%
  \BibitemOpen
  \bibfield  {author} {\bibinfo {author} {\bibfnamefont {L.}~\bibnamefont
  {Pollet}},\ }\bibfield  {title} {\bibinfo {title} {Recent developments in
  quantum monte carlo simulations with applications for cold gases},\ }\href
  {https://doi.org/10.1088/0034-4885/75/9/094501} {\bibfield  {journal}
  {\bibinfo  {journal} {Reports on Progress in Physics}\ }\textbf {\bibinfo
  {volume} {75}},\ \bibinfo {pages} {094501} (\bibinfo {year}
  {2012})}\BibitemShut {NoStop}%
\bibitem [{\citenamefont {Akaturk}\ \emph {et~al.}()\citenamefont {Akaturk},
  \citenamefont {Ezzell},\ and\ \citenamefont {Hen}}]{advancedMeasurements}%
  \BibitemOpen
  \bibfield  {author} {\bibinfo {author} {\bibfnamefont {E.}~\bibnamefont
  {Akaturk}}, \bibinfo {author} {\bibfnamefont {N.}~\bibnamefont {Ezzell}},\
  and\ \bibinfo {author} {\bibfnamefont {I.}~\bibnamefont {Hen}},\ }\bibfield
  {title} {\bibinfo {title} {Permutation matrix representation quantum monte
  carlo: advanced measurement techniques},\ }\bibinfo {note} {(in
  preparation)}\BibitemShut {NoStop}%
\bibitem [{\citenamefont {Raventós}\ \emph {et~al.}(2017)\citenamefont
  {Raventós}, \citenamefont {Graß}, \citenamefont {Lewenstein},\ and\
  \citenamefont {Juliá-Díaz}}]{OBDM}%
  \BibitemOpen
  \bibfield  {author} {\bibinfo {author} {\bibfnamefont {D.}~\bibnamefont
  {Raventós}}, \bibinfo {author} {\bibfnamefont {T.}~\bibnamefont {Graß}},
  \bibinfo {author} {\bibfnamefont {M.}~\bibnamefont {Lewenstein}},\ and\
  \bibinfo {author} {\bibfnamefont {B.}~\bibnamefont {Juliá-Díaz}},\
  }\bibfield  {title} {\bibinfo {title} {Cold bosons in optical lattices: a
  tutorial for exact diagonalization},\ }\href
  {https://doi.org/10.1088/1361-6455/aa68b1} {\bibfield  {journal} {\bibinfo
  {journal} {Journal of Physics B: Atomic, Molecular and Optical Physics}\
  }\textbf {\bibinfo {volume} {50}},\ \bibinfo {pages} {113001} (\bibinfo
  {year} {2017})}\BibitemShut {NoStop}%
\bibitem [{\citenamefont {Penrose}\ and\ \citenamefont
  {Onsager}(1956)}]{penrose1}%
  \BibitemOpen
  \bibfield  {author} {\bibinfo {author} {\bibfnamefont {O.}~\bibnamefont
  {Penrose}}\ and\ \bibinfo {author} {\bibfnamefont {L.}~\bibnamefont
  {Onsager}},\ }\bibfield  {title} {\bibinfo {title} {Bose-einstein
  condensation and liquid helium},\ }\href
  {https://doi.org/10.1103/PhysRev.104.576} {\bibfield  {journal} {\bibinfo
  {journal} {Phys. Rev.}\ }\textbf {\bibinfo {volume} {104}},\ \bibinfo {pages}
  {576} (\bibinfo {year} {1956})}\BibitemShut {NoStop}%
\bibitem [{\citenamefont {Penrose}(1951)}]{penrose2}%
  \BibitemOpen
  \bibfield  {author} {\bibinfo {author} {\bibfnamefont {O.}~\bibnamefont
  {Penrose}},\ }\bibfield  {title} {\bibinfo {title} {Cxxxvi. on the quantum
  mechanics of helium ii},\ }\href {https://doi.org/10.1080/14786445108560954}
  {\bibfield  {journal} {\bibinfo  {journal} {The London, Edinburgh, and Dublin
  Philosophical Magazine and Journal of Science}\ }\textbf {\bibinfo {volume}
  {42}},\ \bibinfo {pages} {1373} (\bibinfo {year} {1951})},\ \Eprint
  {https://arxiv.org/abs/https://doi.org/10.1080/14786445108560954}
  {https://doi.org/10.1080/14786445108560954} \BibitemShut {NoStop}%
\bibitem [{\citenamefont {Leggett}(1998)}]{Leggett}%
  \BibitemOpen
  \bibfield  {author} {\bibinfo {author} {\bibfnamefont {A.~J.}\ \bibnamefont
  {Leggett}},\ }\bibfield  {title} {\bibinfo {title} {On the superfluid
  fraction of an arbitrary many-body system at t=0},\ }\href
  {https://doi.org/10.1023/B:JOSS.0000033170.38619.6c} {\bibfield  {journal}
  {\bibinfo  {journal} {Journal of Statistical Physics}\ }\textbf {\bibinfo
  {volume} {93}},\ \bibinfo {pages} {927} (\bibinfo {year} {1998})}\BibitemShut
  {NoStop}%
\bibitem [{\citenamefont {Pollock}\ and\ \citenamefont
  {Ceperley}(1987)}]{pollock1987path}%
  \BibitemOpen
  \bibfield  {author} {\bibinfo {author} {\bibfnamefont {E.~L.}\ \bibnamefont
  {Pollock}}\ and\ \bibinfo {author} {\bibfnamefont {D.~M.}\ \bibnamefont
  {Ceperley}},\ }\bibfield  {title} {\bibinfo {title} {Path-integral
  computation of superfluid densities},\ }\href@noop {} {\bibfield  {journal}
  {\bibinfo  {journal} {Physical Review B}\ }\textbf {\bibinfo {volume} {36}},\
  \bibinfo {pages} {8343} (\bibinfo {year} {1987})}\BibitemShut {NoStop}%
\bibitem [{\citenamefont {dePina}(1995)}]{depina95}%
  \BibitemOpen
  \bibfield  {author} {\bibinfo {author} {\bibfnamefont {J.~C.}\ \bibnamefont
  {dePina}},\ }\bibfield  {title} {\bibinfo {title} {Applications of shortest
  path methods},\ }\href@noop {} {\  (\bibinfo {year} {1995})}\BibitemShut
  {NoStop}%
\bibitem [{\citenamefont {Kavitha}\ \emph {et~al.}(2008)\citenamefont
  {Kavitha}, \citenamefont {Mehlhorn}, \citenamefont {Michail},\ and\
  \citenamefont {Paluch}}]{Kavitha08}%
  \BibitemOpen
  \bibfield  {author} {\bibinfo {author} {\bibfnamefont {T.}~\bibnamefont
  {Kavitha}}, \bibinfo {author} {\bibfnamefont {K.}~\bibnamefont {Mehlhorn}},
  \bibinfo {author} {\bibfnamefont {D.}~\bibnamefont {Michail}},\ and\ \bibinfo
  {author} {\bibfnamefont {K.~E.}\ \bibnamefont {Paluch}},\ }\bibfield  {title}
  {\bibinfo {title} {An {$\tilde{O}(m^2 n)\ $}algorithm for minimum cycle basis
  of graphs},\ }\href@noop {} {\bibfield  {journal} {\bibinfo  {journal}
  {Algorithmica}\ }\textbf {\bibinfo {volume} {52}},\ \bibinfo {pages} {333}
  (\bibinfo {year} {2008})}\BibitemShut {NoStop}%
\end{thebibliography}%

\appendix

\section{The off-diagonal partition function expansion\label{apx:odp}}

Here, we describe the expansion of the partition function in terms of the off-diagonal operators of the Hamiltonian. The partition function is given as:
\begin{equation}
Z = Tr[e^{-\beta H}]
\end{equation}
Replace trace by explicit sum $\sum \langle z\vert\cdot\vert z\rangle$, then expand the exponent in Taylor series in $\beta$
\begin{equation}
\begin{aligned}
Z &= \sum_{z} \sum_{n=0}^{\infty} \frac{\beta^n}{n!} \langle z \vert (-H)^n \vert z \rangle \\
&= \sum_{z} \sum_{n=0}^{\infty} \frac{\beta^n}{n!} \langle z \vert \ \bigg(1 -D_0 - \sum_{j=1} D_j P_j \bigg)^n \ \vert z \rangle \\
&= \sum_{z} \sum_{n=0}^{\infty} \sum_{\{ S_{\mathbf{i}_n} \}} \frac{\beta^n}{n!} \langle z \vert S_{\mathbf{i}_n} \vert z \rangle
\end{aligned}
\end{equation}
in last step, $(-H)^n$ expressed in all sequences of length $n$ composed of products of $D_0$ and $D_j P_j$ which is denoted as $\{ S_{ \mathbf{i}_n } \}$, $\mathbf{i}_n = (i_1,i_2,\ldots,i_n)$, $i_j \in \{ 0,\ldots,M \}$ $j \in \{ 1,\ldots,n \}$
\begin{equation}
\begin{aligned}
Z &= \sum_{z} \sum_{q=0}^{\infty} \sum_{\{ S_{q} \}} \bigg( \prod_{j=1}^{q} d_{z_j}^{(i_j)} \bigg) \langle z \vert S_{\mathbf{i}_n} \vert z \rangle \\ &\bigg( \sum_{n=q}^{\infty} \frac{\beta^n (-1)^n}{n!} \times \sum_{\sum k_i = n-q} (E_{z_0})^{k_0} \cdot \ldots \cdot (E_{z_q})^{k_q} \bigg)
\end{aligned}
\end{equation}
where $E_{z_i} = \langle z_i \vert D_0 \vert z_i \rangle$.
\begin{equation}
d_{z_j}^{(i_j)} = \langle z_j \vert D_{i_j} \vert z_j \rangle
\end{equation}
$S_{\mathbf{i}_q} = P_{i_q} \ldots P_{i_2} P_{i_1}, \vert z_0 \rangle = \vert z \rangle, P_{i_j} \vert z_j \rangle = \vert z_{j+1} \rangle $. $\vert z_j \rangle = \vert z_{(i_1,i_2,\ldots,i_j)} \rangle$
$n \rightarrow n+q$ gives:
\begin{equation}
\begin{aligned}
Z &= \sum_{z} \sum_{q=0}^{\infty} \sum_{\{ S_{q} \}} \langle z \vert S_{\mathbf{i}_q} \vert z \rangle \Bigg( (-\beta)^q \bigg( \prod_{j=1}^{q} d_{z_j}^{(i_j)} \bigg) \\ &\times \sum_{n=0}^{\infty} \frac{{-\beta}^n}{(n+q)!} \sum_{\sum ki=n} (E_{z_0})^{k_0} \cdot \ldots \cdot (E_{z_q})^{k_q} \Bigg)
\end{aligned}
\end{equation}
$\{ E_{z_i} \}$ are classical energies of $\vert z_i \rangle$ which are created by the application of $S_{{\bf i}_q}$.
\begin{equation}
\begin{aligned}
Z &= \sum_{z} \sum_{q=0}^{\infty} \bigg( \prod_{j=1}^{q} d_{z_j}^{(i_j)} \bigg) \sum_{ \{ S_q \} } \langle z \vert S_{\mathbf{i}_q} \vert z \rangle \\ &\times \Bigg( \sum_{ \{ k_i \} = ( 0, \ldots, 0 ) }^{ ( \infty, \ldots, \infty ) } \frac{{-\beta}^q}{(q+\sum k_i)!} \prod_{j=0}^{q} (-\beta E_{z_{z_j}})^{k_j} \Bigg)
\end{aligned}
\end{equation}
\begin{equation}
\sum_{ \{ k_i \} } \frac{{-\beta}^q}{(q+\sum k_i)!} \prod_{j=0}^{q} (-\beta E_{z_{z_j}})^{k_j} = e^{-\beta[ E_{z_0}, \ldots, E_{z_q}]}
\end{equation}
$[ E_{z_0}, \ldots, E_{z_q}]$ is a multiset of energies
\begin{equation}
F[ E_{z_0}, \ldots, E_{z_q} ] \equiv \sum_{j=0}^{q} \frac{ F(E_{z_j}) }{ \prod_{k \neq j} ( E_{z_j} - E_{z_k} ) }
\end{equation}
$F$ is called divided differences, defined for real valued variables $[ E_{z_0}, \ldots, E_{z_q}]$.
\begin{equation}
Z = \sum_{z} \sum_{q=0}^{\infty} \sum_{\{ S_{q} \}} \langle z \vert S_{\mathbf{i}_q} \vert z \rangle D_{ ( z, S_{\mathbf{i}_q} ) } e^{-\beta[ E_{z_0}, \ldots, E_{z_q}]}
\end{equation}
where
\begin{equation}
D_{ ( z, S_{\mathbf{i}_q} ) } = \prod_{j=1}^{q} d_{z_j}^{(i_j)}
\end{equation}
Note that, expansion of $Z$ is not an expansion in $\beta$. It begins with a Taylor series expansion in $\beta$ but regrouping of terms into the exponent of divided-differences means no longer a high temperature expansion. \note{check org work Page4Column2}

One can interpret $Z$ expansion as a sum of weights. $Z = \sum_{\{ {\cal C} \}} W_{\cal C}$, where $\{ {\cal C} \}$ is all distinct pairs $\{ \vert z \rangle, S_{ \mathbf{i}_q } \}$
\begin{equation}
W_{\cal C} = D_{(z, S_{\mathbf{i}_q})}e^{-\beta[ E_{z_0}, \ldots, E_{z_q}]}
\end{equation}
$W_{\cal C}$ is the configuration weight. 
$\langle z \vert S_{\mathbf{i}_q} \vert z \rangle$ evaluates to either $1$ or $0$. Since $P_j, j \neq 0$ has no fixed points, $S_{\mathbf{i}_q} = 1$ implies $S_{\mathbf{i}_q} = \identity$. Then,
\begin{equation}
Z = \sum_{z} \sum_{ S_{ \mathbf{i}_q } = \mathds{1} } D_{( z, S_{ \mathbf{i}_q } )}e^{-\beta[ E_{z_0}, \ldots, E_{z_q}]} \,.
\end{equation}

\end{document}